\documentclass[paper = a4,  % Die Papiergröße wird bei KOMA-Klasse automatisch
        pagesize = auto,    % Setzen des Seitenformates in der Ausgabedatei
        twoside=true,       % Dokument wird zweiseitig gesetzt für true; für
        titlepage = false,
        abstract = true,
]{scrartcl}

\usepackage[utf8]{inputenc}
\usepackage[T1]{fontenc}
\usepackage[english]{babel}
\usepackage{lmodern}
\usepackage[noblocks]{authblk}

\usepackage[style=authoryear,backend=biber,isbn=false,doi=true]{biblatex}
\usepackage{csquotes}
\usepackage{amsmath}
\allowdisplaybreaks[1]     % Seitenumbrueche in Gleichungen erlauben (kein
\usepackage{amssymb}
\usepackage{amsthm}
\usepackage{mathtools}
\DeclareMathOperator*{\argmin}{argmin}

\usepackage[export]{adjustbox}
\usepackage{subfig}
\usepackage[english, ruled, vlined]{algorithm2e}
\SetAlCapSty{textnormal}        %keine fette Bildunterschrift fuer Algorithmen
\usepackage{booktabs}
\usepackage{url}
\usepackage{siunitx}
\DeclareSIUnit[number-unit-product = {}]{\dt}{dt}
\DeclareSIUnit[number-unit-product = {}]{\year}{yr}
\DeclareSIUnit[number-unit-product = {}]{\Phosphate}{P}
\DeclareSIUnit[number-unit-product = {}]{\Timestep}{dt}
\usepackage{pdfpages}

\usepackage[%bookmarksopen = true,
pdftitle={Surrogate-based optimization using an artificial neural network for a parameter identification in a 3D marine ecosystem model},%
pdfauthor={Markus Pfeil and Thomas Slawig}]{hyperref}

\title{Surrogate-based optimization using an artificial neural network for a parameter identification in a 3D marine ecosystem model}

\author[1]{Markus Pfeil}
\author[1]{Thomas Slawig}
\affil[1]{Kiel Marine Science (KMS) - Centre for Interdisciplinary Marine Science, Dep. of Computer Science, Kiel University, 24098 Kiel, Germany (\{mpf, ts\}@informatik.uni-kiel.de).}

\date{\vspace{-5ex}}

\begin{document}

  \maketitle

  \begin{abstract}
Parameter identification for marine ecosystem models is important for the
assessment and validation of marine ecosystem models against observational data.
The surrogate-based optimization (SBO) is a computationally efficient method to
optimize complex models. SBO replaces the computationally expensive
(high-fidelity) model by a surrogate constructed from a less accurate but
computationally cheaper (low-fidelity) model in combination with an appropriate
correction approach, which improves the accuracy of the low-fidelity model. To
construct a computationally cheap low-fidelity model, we tested three different
approaches to compute an approximation of the annual periodic solution (i.e.,
a steady annual cycle) of a marine ecosystem model: firstly, a reduced number of
spin-up iterations (several decades instead of millennia), secondly, an
artificial neural network (ANN) approximating the steady annual cycle and,
finally, a combination of both approaches. Except for the low-fidelity model
using only the ANN, the SBO yielded a solution close to the target and reduced
the computational effort significantly. If an ANN approximating appropriately a
marine ecosystem model is available, the SBO using this ANN as low-fidelity
model presents a promising and computational efficient method for the
validation.
\end{abstract}

  \section{Introduction}
  \label{sec:Introduction}

    In the field of climate research, marine ecosystem models are part and parcel
of the analysis of changes in the marine ecosystem influenced by diverse
biogeochemical processes. As part of the global carbon cycle, the ocean takes
up, for example, $\textrm{CO}_{2}$ from the atmosphere and stores it. A marine
ecosystem model consists of a global circulation model coupled with a
biogeochemical model considering the interactions of the physical and
biogeochemical processes \parencite[cf.][]{Fasham03, SarGru06, FenNeu04}. In
particular, the equations and variables describing the physical processes are
well known. Conversely, there is generally no set of equations and variables to
describe the biogeochemical processes which is why many biogeochemical models
exist that differ in their complexity by the number of state variables and
parametrizations \parencite[see e.g.,][]{KrKhOs10}. Therefore, validation and
assessment of the biogeochemical models, which include a parameter optimization
and a discussion of simulation results, are necessary by an evaluation of the
model outputs against observational data \parencite{FLSW01}.

Parameter identification to validate biogeochemical models, especially for
three-di\-men\-sion\-al models, is a challenging task with a large
computationally effort. The parameter identification determines or adjusts the
model parameters to fit the model to existing measurement data or a corresponding
model output \parencite{BanKun89}. For marine ecosystem models, this
corresponds to the solving of a nonlinear optimization problem that typically
requires many evaluations of the model-data misfit function, also called cost or
objective function. As a consequence, such optimizations using conventional
optimization algorithms are very time-consuming or infeasible even on
high-performance computers because a single model evaluation already is
computationally expensive \parencite{Osc06}. To date, the search for a feasible
optimization for global marine ecosystem models has included, for instance,
approaches of a statistical emulator technique as well as a gradient-free
method \parencite{KwoPri06, KwoPri08, MaFeDo12}. Due to the high computational
effort, the acceleration of the optimization process itself or of the
underlying simulation is still of great interest.

\emph{Surrogate-based optimization} (SBO) is a computationally efficient method
to optimize complex models (so-called \emph{high-fidelity models}), whose
simple model evaluation already requires enormous computational effort
\parencite{QHSGVT05, ForKea09, BCDMBM04, LeiKoz10}. The surrogate, which is
computationally cheap and a reasonable approximation of the high-fidelity
model, replaces the high-fidelity model in the optimization process.
Consequently, the SBO estimates the optimal parameters of the high-fidelity
model by optimizing the surrogate. There are several variants to construct the
surrogate, such as function-approximation surrogates \parencite{QHSGVT05,
SPKA01, SmoSch04} or physics-based ones \parencite{Son03} using a
\emph{low-fidelity model} that is a less accurate approximation of the
high-fidelity model. The latter approach is used in this paper because coarser
discretizations (in time and/or space) are common in climate research
\parencite{McgHen14}. The application of SBO for the parameter identification of
marine ecosystem models has already reduced computational costs compared to
conventional optimizations because the number of high-fidelity model evaluations
were much smaller \parencite{PPKOS13, PrKoSl11, PrKoSl13}.

Several strategies enable the reduction of the computational effort simulating
marine ecosystem models. The fully coupled simulation of a marine ecosystem
model to compute a steady annual cycle is computationally expensive because it
requires a long-time integration over several millennia, but a single evaluation
is already computational expensive \parencite[cf.][]{BeDiWu08, Bryan84,
DaMcLa96, WunHei08, SibWun11, Osc06}. As opposed to the fully coupled simulation
(the so-called \emph{online} simulation), the \emph{offline} simulation neglects
the impact of the biogeochemical model on the ocean circulation to reduce the
computational effort using pre-computed data of the ocean currents. Moreover,
the \emph{transport matrix method} (TMM) reduces the computation of the global
ocean circulation to a matrix-vector multiplication and, thus, decouples the
evaluation of the biogeochemical model from the ocean circulation
\parencite{KhViCa05, Kha07}. This approach also lowers the computational effort
with a tolerable loss of accuracy. These two strategies are only two examples
to reduce the computational effort.

Deep Learning enables predictions of steady annual cycles for a marine
ecosystem model using the features learned from other steady annual cycles. In
recent years, Deep Learning \parencite{GoBeCo16, LeBeHi15} based on
\emph{artificial neural networks} (ANNs) allowed breakthroughs in many areas,
such as classification, speech recognition, computer vision or bioinformatics.
The power of Deep Learning lies in recognizing patterns in data provided to
train the neural network and predicting corresponding data based on the learned
patterns. We applied Deep Learning to predict steady annual cycles from model
parameters so that the long-time integration was not longer necessary
\parencite{PfeSla21}. 

The combination of a low-fidelity model and correction technique affects both
the accuracy and the computational effort of the surrogate that was used for the
SBO \parencite[cf.][]{Son03, PPKOS13}. The low-fidelity model is always a
compromise between the accuracy and the computational effort of the
high-fidelity model approximation. An ANN offers new opportunities to
approximate a steady annual cycle for the use as low-fidelity model. While the
accuracy using only the ANN as low-fidelity model was not sufficient for the
SBO, the SBO which used a physics-based low-fidelity model improved by the ANN
showed good results. In the present paper, we applied synthetic data generated
by the high-fidelity model to assess the feasibility of the SBO with the
different low-fidelity models. Instead of real measurement data, the use of
synthetic data avoided uncertainties, such as the required structural complexity
of the model to reconstruct real measurement data, errors in the measurements
and the performance of the optimization itself.

This paper is structured as follows: after an introduction into marine
ecosystem models including the computation of steady annual cycles using the
TMM in Section \ref{sec:Model}, Section \ref{sec:ArtificialNeuralNetwork}
contains the description of the methods used to train an ANN together with the
resulting ANN. In Section \ref{sec:SurrogateBasedOptimization}, we describe the
SBO. Numerical results of the SBO using different low-fidelity models are
presented in Section \ref{sec:Results}. The paper closes with a summary and
conclusions.

  \section{Model description}
  \label{sec:Model}

    A marine ecosystem model describes the interplay between the ocean circulation
and the interactions and biogeochemistry among ocean biota. The marine ecosystem
is represented by a given number of ecosystem species (or tracers) which are
substances in the ocean water and subject to chemical or biochemical reactions.
In the marine ecosystem model, both the ocean circulation affects the tracer
concentrations and, vice versa, the tracers affect the ocean circulation. A
fully coupled model (\emph{online} model) includes both of these interactions
and, hence, simulating such a model is computationally expensive and often
limited to single model evaluations \parencite{Osc06}. In contrast, an
\emph{offline} model simplifies the complexity by neglecting the influence of
the tracers on the ocean circulation. As a result of this one-way coupling, we
can use a pre-computed ocean circulation for the simulation.

\subsection{Model equations for marine ecosystems}
\label{sec:ModelEquation}

  A system of partial differential equations models the marine ecosystem. The
  complexity of the marine ecosystem models varies, inter alia, by the number
  of tracers, which defines the size of the system of differential equations.
  The differential equations in the system are of the Lotka-Volterra or
  predator-prey type \parencite{Lot10, Vol31}. In the rest of this paper, we
  consider marine ecosystem models using an offline model and $n_y \in
  \mathbb{N}$ tracers on a spatial domain $\Omega \subset \mathbb{R}^3$ (i.e.,
  the ocean) and a time interval $[0,1]$ (i.e., one model year). The mapping
  $y_i: \Omega \times [0,1] \rightarrow \mathbb{R}$, $i \in \left\{1, \ldots,
  n_y \right\}$, describes the tracer concentration of the single tracer $y_i$
  and $\mathbf{y} := \left( y_i \right)_{i=1}^{n_{y}}$ summarizes the tracer
  concentrations of all tracers. The system of parabolic partial differential
  equations
  \begin{align}
  \label{eqn:Modelequation}
    %\begin{split}
      \frac{\partial y_i}{\partial t} (x,t)
           + \left( D (x,t) + A(x,t) \right) y_i (x,t)
        &= q_i \left( x, t, \mathbf{y}, \mathbf{u} \right),
        & x \in \Omega, t &\in [0,1], \\
    \label{eqn:Boundarycondition}
      \frac{\partial y_i}{\partial n} (x,t) &= 0,
        & x \in \partial \Omega, t &\in [0,1],
   % \end{split}
  \end{align}
  for $i = 1, \ldots, n_{y}$, defines the tracer transport of a marine
  ecosystem model. The linear operators $A: \Omega \times [0,1] \rightarrow
  \mathbb{R}$ and $D: \Omega \times [0,1] \rightarrow \mathbb{R}$ correspond
  to the advection and diffusion coming from the ocean circulation while the
  term $q_i: \Omega \times [0,1] \rightarrow \mathbb{R}$ summarizes all
  biogeochemical terms for the tracer $y_i$. The homogeneous Neumann boundary
  condition \eqref{eqn:Boundarycondition} includes the normal derivative
  describing an absence of fluxes on the boundary.
  
  The spatial tracer transport in marine water depends on the ocean currents
  in form of spatially discretized advection and diffusion. The advection
  including a given velocity field $v: \Omega \times [0,1] \rightarrow
  \mathbb{R}^3$ is modeled as
  \begin{align}
    \label{eqn:Advection}
    A(x,t) y_i (x,t) &:= \textrm{div} \left( v(x,t) y_i (x,t) \right),
    & x \in \Omega, t &\in [0,1]
  \end{align}
  for $i \in \{1, \ldots, n_y\}$. The diffusion models the turbulent effects
  of the ocean circulation. The molecular diffusion of the tracers themselves,
  conversely, is neglected because this diffusion is much smaller than the
  diffusion induced by turbulence. In ocean circulation modeling, the quite
  different scales in horizontal and vertical direction requires a splitting
  $D = D_h + D_v$ and an implicit treatment of the vertical part in the time
  integration. Both directions of the diffusion are modeled in the
  second-order form as
  \begin{align}
    \label{eqn:Diffusion-horizontal}
    D_{h} (x,t) y_i (x,t) &:= -{\textrm{div}_h} \left( \kappa_h (x,t) \nabla_h
                                y_i (x,t) \right)
    & x \in \Omega, t &\in [0,1],\\
    \label{eqn:Diffusion-vertical}
    D_{v} (x,t) y_i (x,t) &:= -\frac{\partial}{\partial z}
            \left(\kappa_{v} (x,t)\frac{\partial y_i}{\partial z}(x,t)\right),
    & x \in \Omega, t &\in [0,1]
  \end{align}
  for $i \in \{1, \ldots, n_y\}$, where $\textrm{div}_h$ and $\nabla_h$ denote
  the horizontal divergence and gradient, $\kappa_h, \kappa_v: \Omega \times
  [0,1] \rightarrow \mathbb{R}$ the diffusion coefficient fields and $z$ is
  the vertical coordinate. Due to the neglect of the molecular diffusion, the
  diffusion coefficients are identical for all tracers.
  
  The biogeochemical model summarizes the biogeochemical processes
  modeled in the marine ecosystem. The nonlinear function $q_i: \Omega \times
  [0,1] \rightarrow~\mathbb{R}, \left( x, t \right) \mapsto q_i \left( x, t,
  \mathbf{y}, \mathbf{u} \right)$ describes the biogeochemical processes for
  the tracer $y_i$, $i \in \{1, \ldots, n_y\}$. More specifically, the
  biogeochemical processes depend, firstly, on the variability of the solar
  radiation in space and time, secondly, on the coupling to other tracers and,
  thirdly, on $n_u \in \mathbb{N}$ model parameters $\mathbf{u} \in
  \mathbb{R}^{n_u}$ (such as growth, loss and mortality rates or sinking
  speed). The biogeochemical model  $\mathbf{q} = \left( q_i
  \right)_{i=1}^{n_{y}}$ combines the biogeochemical processes for all tracers.
  In contrast to the biogeochemical model, the marine ecosystem model, moreover,
  includes the effects of the ocean circulation and, thus, comprises the whole
  system \eqref{eqn:Modelequation} to \eqref{eqn:PeriodicCondition}.
  
  For the marine ecosystem model, an annual periodic solution of
  \eqref{eqn:Modelequation} and \eqref{eqn:Boundarycondition} (i.e., a steady
  annual cycle) fulfills
  \begin{align}
   \label{eqn:PeriodicCondition}
    y_i (x, 0) &= y_i (x, 1), & x &\in \Omega,
  \end{align}
  for $i \in \{1, \ldots, n_y\}$. For this purpose, we assume that the
  operators $D, A$ and the functions $q_i$ also are annually periodic in time.

\subsection{Biogeochemical models}
\label{sec:BiogeochemicalModels}

  We applied two global biogeochemical models with increased complexity,
  the N and N-DOP model \parencite{KrKhOs10, PiwSla16}. In the following, we
  briefly introduce these models. For a detailed description of the modeled
  processes and model equations, we refer to \textcite{KrKhOs10} and
  \textcite{PiwSla16}.

  The light intensity affects the biogeochemical processes in the ocean. The
  light limitation function $I: \Omega \times [0, 1] \rightarrow
  \mathbb{R}_{\geq 0}$ models the light intensity depending on the insolation.
  This is based on the astronomical formula of \textcite{PalPla76} and
  considers the ice cover as well as the exponential attenuation of water.
  According to the light intensity, the ocean is divided into a euphotic (sun
  lit) zone of about \SI{100}{\metre} and an aphotic zone below. A fast and
  dynamic turnover of phosphorus, especially, is part of the euphotic zone, for
  instance via photosynthesis, grazing or mortality. Furthermore, a part of
  the biological production sinks from the euphotic zone as a particulate matter
  to depth and remineralizes there according to the empirical power-law
  relationship \parencite{MKKB87}.
  
  %table:ModelParameter
  \begin{table}[tb]
    \centering
    \caption{Model parameters of the biogeochemical models.}
    \label{table:ModelParameter}
    \begin{tabular}{l l l}
      \hline
      Parameter & Description & Unit \\
      \hline
      $k_w$ & Attenuation coefficient of water & \si{\per \metre} \\
      $\mu_P$ & Maximum growth rate & \si{\per \day} \\
      $K_N$ & Half saturation constant for \si{PO_4} uptake & \si{\milli \mol \Phosphate \per \cubic \metre} \\
      $K_I$ & Light intensity compensation & \si{\watt \per \square \metre} \\
      $\sigma_\text{DOP}$ & Fraction of phytoplankton losses assigned to DOP & \si{1} \\
      $\lambda'_\text{DOP}$ & Decay rate & \si{\per \year} \\
      $b$ & Implicit representation of sinking speed & \si{1}  \\
      \hline
    \end{tabular}
  \end{table}
  
  The N model contains only one tracer describing phosphate ($\textrm{PO}_4$)
  as inorganic nutrients \parencite[cf.][]{BacMai90, KrKhOs10}. We denote this
  model as N model (\textrm{N} for nutrients), i.e., $\mathbf{y} =
  (\mathbf{y}_{\textrm{N}})$. The $n_u = 5$ model parameters $\mathbf{u} =
  \left( k_w, \mu_P, K_N, K_I, b \right)$ (see Table
  \ref{table:ModelParameter}) control the biogeochemical processes. The
  phytoplankton production (or biological uptake) is defined as
  \begin{align}
    \label{eqn:Phytoplankton}
    f_P: \Omega \times [0,1] \rightarrow \mathbb{R},
      f_P (x, t) &= \mu_P y_P^* \frac{I(x,t)}{K_I + I(x,t)}
                    \frac{\mathbf{y}_N (x,t)}{K_N + \mathbf{y}_N (x,t)}
  \end{align}
  and depends on available nutrients and light. Moreover, the uptake of
  nutrients by phytoplankton is limited using a half saturation function, and
  applies a model parameter $\mu_P$ for the maximum production rate as well as
  a prescribed concentration of phytoplankton
  $y_P^* = 0.0028$~\si{\milli\mole\Phosphate\per\cubic\metre}.

  Additional to nutrients (\textrm{N}), the N-DOP model includes dissolved
  organic phosphorus (\textrm{DOP}) \parencite[cf.][]{BacMai91, PaFoBo05,
  KrKhOs10}, i.e., $\mathbf{y} = ( \mathbf{y}_{\textrm{N}},
  \mathbf{y}_{\textrm{DOP}} )$. Using the same phytoplankton production
  \eqref{eqn:Phytoplankton} as the N model, the N-DOP contains $n_u = 7$ model
  parameters $\mathbf{u} = \left( k_w, \mu_P, K_N, K_I, \sigma_{\text{DOP}},
  \lambda'_{\text{DOP}}, b \right)$ (see Table \ref{table:ModelParameter}) to
  describe the internal processes.

\subsection{Transport matrix method}
\label{sec:TransportMatrixMethod}

  The \emph{transport matrix method} (TMM) efficiently approximates the tracer
  transport by matrix-vector multiplications \parencite{KhViCa05, Kha07}.
  Instead of implementing directly a discretization scheme for the advection
  and diffusion operators $A$ and $D$, the TMM approximates the ocean
  circulation by matrices because the application of the two operators on a
  spatially discretized tracer vector is linear and, therefore, the
  discretized advection-diffusion equation can be written as a linear matrix
  equation. For this purpose, the TMM computes and stores the matrices
  obtained by applying the discretized operators on a discrete tracer vector.
  Consequently, the matrices contain the transport of all parameterized
  processes represented in the underlying ocean circulation model
  \parencite{KhViCa05}.

  The TMM reduces a time step of the simulation of a marine ecosystem model to
  matrix-vector multiplications and an evaluation of the biogeochemical model.
  We assume that, firstly, the grid with $n_x \in \mathbb{N}$ grid points
  $\left( x_k \right)_{k=1}^{n_x}$ is a spatial discretization of the domain
  $\Omega$ (i.e., the ocean) and, secondly, the time steps $t_0, \ldots,
  t_{n_{t}} \in [0,1]$, $n_t \in \mathbb{N}$, specified by
  \begin{align*}
    t_j &:= j \Delta t, & j &= 0, \ldots, n_t, & \Delta t &:= \frac{1}{n_t},
  \end{align*}
  define an equidistant grid of the time interval $[0,1]$  (i.e., one model
  year). For the time instant $t_j$, $j \in \{0, \ldots, n_t - 1\}$, the
  vector $\mathbf{y}_{j} := \left( \mathbf{y}_{ji} \right)_{i=1}^{n_y} \in
  \mathbb{R}^{n_y n_x}$ combines the numerical approximations for all tracers
  at time instant $t_j$ using a reasonable concatenation whereby vector
  \begin{align*}
    \mathbf{y}_{ji} &\approx \left( y_{i} \left( t_{j}, x_{k} \right)
                         \right)_{k=1}^{n_x} \in \mathbb{R}^{n_x}
  \end{align*}
  denotes the approximation of the spatially discrete tracer $y_i$, $i \in
  \{1, \ldots, n_y\}$. Analogously,
  \begin{align*}
    \mathbf{q}_{ji} &\approx \left( q_i \left( x_k, t_j, \mathbf{y_j}, \mathbf{u} \right)
                     \right)_{k=1}^{n_x} \in \mathbb{R}^{n_x}
  \end{align*}
  contains the spatially discretized biogeochemical term of tracer $y_i$, $i
  \in \{1, \ldots, n_y\}$, at time instant $t_j$, $j \in \{0, \ldots, n_t -
  1\}$, and $\mathbf{q}_j := \left( \mathbf{q}_{ji} \right)_{i=1}^{n_y}$
  summarizes this for all tracers at time instant $t_j$. In equation
  \eqref{eqn:Modelequation}, we discretize the advection and the horizontal
  diffusion explicitly and the vertical diffusion implicitly. Using the
  spatially discrete counterparts $\mathbf{A}_j, \mathbf{D}_j^h$ and
  $\mathbf{D}_{j}^v$ of the operators $A, D_h$ and $D_v$ at time instant
  $t_j$, $j \in \{0, \ldots, n_t -1\}$, the semi-implicit Euler scheme of
  \eqref{eqn:Modelequation} results in a time-stepping
  \begin{align*}
    \mathbf{y}_{j+1} &=  \left( \mathbf{I} + \Delta t \mathbf{A}_j
                            + \Delta t \mathbf{D}_j^h \right) \mathbf{y}_j
                     + \Delta t \mathbf{D}_j^v \mathbf{y}_{j+1}
                     + \Delta t \mathbf{q}_j \left( \mathbf{y}_j, \mathbf{u} \right),
        & j &= 0, \ldots, n_t -1
  \end{align*}
  with the identity matrix $\mathbf{I} \in \mathbb{R}^{n_x \times n_x}$.
  Defining the explicit and implicit transport matrices
  \begin{align*}
    \mathbf{T}_{j}^{\text{exp}} &:= \mathbf{I} + \Delta t \mathbf{A}_j
                                    + \Delta t \mathbf{D}_j^h \in \mathbb{R}^{n_x \times n_x}, \\
    \mathbf{T}_{j}^{\text{imp}} &:= \left( \mathbf{I} - \Delta t \mathbf{D}_j^v
                                    \right)^{-1} \in \mathbb{R}^{n_x \times n_x}
  \end{align*}
  for each time instant $t_j$, $j \in \{0, \ldots, n_t - 1\}$, we obtain for a
  time step of the marine ecosystem model using the TMM
  \begin{align}
    \label{eqn:TMM}
    \mathbf{y}_{j+1} &= \mathbf{T}_{j}^{\text{imp}}
                     \left( \mathbf{T}_{j}^{\text{exp}} \mathbf{y}_j
                        + \Delta t \mathbf{q}_j \left( \mathbf{y}_j, \mathbf{u} \right)
                        \right)
                 =: \varphi_j \left( \mathbf{y}_j, \mathbf{u} \right),
        & j &= 0, \ldots, n_t - 1.
  \end{align}

  The transport matrices are sparse representations of the monthly averaged
  tracer transport. The explicit and implicit transport matrices are sparse
  because these matrices are generated with a grid-point based ocean
  circulation model and the implicit ones (i.e., the inverse of the
  discretization matrices) contain only the vertical part of the diffusion. In
  practical computations, monthly averaged matrices are stored and
  interpolated linearly to compute an approximation for any time instant
  $t_j$, $j=0, \ldots, n_t -1$, because storing the matrices for all
  time-steps in a year is practically impossible. Assuming annually
  periodicity of the ocean circulation, the TMM approximates the ocean
  circulation using twelve pairs of pre-computed transport matrices. In the
  present paper, these matrices are computed with the MIT ocean model
  \parencite{MAHPH97} using a global configuration with a latitudinal and
  longitudinal resolution of \ang{2.8125} and \num{15} vertical layers.

\subsection{Computation of steady annual cycles}
\label{sec:ComputationSteadyAnnualCycles}

  A periodic solution (i.e., a steady annual cycle) in a fully discrete
  setting fulfills 
  \begin{align*}
    \mathbf{y}_{n_t} &= \mathbf{y}_0
  \end{align*}
  applying the above iteration \eqref{eqn:TMM} over one year model time. The
  nonlinear mapping
  \begin{align*}
    \Phi &:= \varphi_{n_t -1} \circ \ldots \circ \varphi_{0}
  \end{align*}
  with $\varphi_j$ specified in \eqref{eqn:TMM} defines the time integration
  of \eqref{eqn:TMM} over one model year. For a marine ecosystem model, the
  steady annual cycle, in particular, is a fixed-point of this mapping.
  Starting from an arbitrary vector $\mathbf{y}^{0} \in \mathbb{R}^{n_y n_x}$
  and model parameters $\mathbf{u} \in \mathbb{R}^{n_u}$, a classical
  fixed-point iteration takes the form
  \begin{align}
    \label{eqn:Spin-upIteration}
    \mathbf{y}^{\ell + 1} &= \Phi \left( \mathbf{y}^{\ell}, \mathbf{u} \right),
                          & \ell = 0, 1, \ldots.
  \end{align}
  The vector $\mathbf{y}^{\ell} \in \mathbb{R}^{n_y n_x}$ contains the tracer
  concentrations at the first time instant of the model year $\ell \in
  \mathbb{N}$ if we interpret the fixed-point iteration
  \eqref{eqn:Spin-upIteration} as pseudo-time stepping or \emph{spin-up}, a
  term which is widely used in ocean and climate research.
  
  As high-fidelity model, we used $n_{f, \ell} := 3000$ model years to compute
  the steady annual cycle with the spin-up. The accuracy of the spin-up is
  already satisfactory after \num{3000} model years \parencite{PPKOS13,
  KrKhOs10}. Nevertheless, the spin-up over more than \num{3000} model years
  can improve the accuracy. The vector 
  \begin{align*}
    \mathbf{y}_f \approx \left( \mathbf{y}^{\ell}_{k} \right)_{k=0, \ldots,
                          n_t -1} \in \mathbb{R}^{n_t n_y n_x}
  \end{align*}
  contains the trajectory of the steady annual cycle computed with the
  high-fidelity model and includes the different tracer concentrations
  $\mathbf{y}^{\ell}_{k} := \varphi_{k-1} \circ \ldots \circ \varphi_{0}
  \left( \mathbf{y}^{\ell}, \mathbf{u} \right)$, for $k \in \{1, \ldots,
  n_t - 1\}$, and $\mathbf{y}^{\ell}_0 := \mathbf{y}^{\ell}$ for each time
  step in the model year.

  \section{The artificial neural network}
  \label{sec:ArtificialNeuralNetwork}

    We applied the prediction of an ANN as an approximation of the steady annual
cycle of the N model \parencite{PfeSla21}. For this purpose, we used a
multilayer perceptron with a feed-forward architecture consisting of an input
layer, three hidden layers and an output layer. More specifically, we employed
a fully connected network, i.e., there do exist all possible connections between
the neurons in two consecutive layers.

The ANN predicted the tracer concentration for the first time of the model year
using the $n_u = 5$ model parameters as input. Therefore, the input layer
consisted of \num{5} neurons, while the output layer contained \num{52749}
neurons. A prediction of the steady annual cycle taking the whole trajectory
into account was not possible because the output layer would consist of almost
\num{2.4} million neurons so that the training of the ANN would be infeasible.
The fully connected network consisted of three hidden layers with \num{10},
\num{25} and \num{211} neurons, respectively.

We trained the ANN with supervised learning \parencite{HaTiFr09} using
backpropagation and used the stochastic gradient descent \parencite{RuHiWi86,
BotBou08} for the optimization of the network weights. The loss function was
the mean squared error. For the neurons of the hidden layers, we applied the 
exponential linear unit \parencite{ClUnHo15} as activation function and the
scaled exponential linear unit \parencite{KUMH17} for the neurons of the output
layer.

The total mass of the predicted tracer concentration was essential for a 
reasonable prediction because the marine ecosystem model preserves mass. We
adjusted the prediction $\mathbf{y}_{\text{ANN}} \in \mathbb{R}^{n_x}$ of the
ANN pointwise by
\begin{align*}
  \tilde{\mathbf{y}}_{\text{ANN}} &= \frac{\sum_{k=1}^{n_x} \left| V_k \right|m }
                                        {\sum_{k=1}^{n_x} \left| V_k \right| \mathbf{y}_{\text{ANN}} (x_k, t_0)}
                                    \, \mathbf{y}_{\text{ANN}}
\end{align*}
using the box volumes $\left| V_k \right|$, $k = 1, \ldots, n_x$, of the
discretization and a global mean tracer concentration $m \in \mathbb{R}_{>0}$
in order to obtain the required overall mass.

A given set of pairs of input and output data is divided into \emph{training},
\emph{validation} and \emph{test data} for the training procedure (i.e., the
optimization of the internal weights of the ANN) using supervised learning. The
optimization adjusts the internal weights by the use of the training data and
monitors the behavior of the already partially optimized network on the
validation data. The test data are used to assess the quality of the trained
network after finishing the training procedure.

  \section{Surrogate-based optimization}
  \label{sec:SurrogateBasedOptimization}

    The SBO \parencite{BCDMBM04, QHSGVT05, ForKea09, LeiKoz10} replaces the high-
fidelity model by a surrogate during the optimization. For many nonlinear
optimization problems, the high computational effort to evaluate the objective
function is a major bottleneck. Conversely, the SBO uses the surrogate to
reduce the computational effort because the evaluation of the surrogate is
computationally much cheaper and still represents a reasonable approximation
of the high-fidelity model. In order to obtain optimal model parameters close
to the high-fidelity optimum, the SBO keeps the surrogate close to the
high-fidelity model by iterative updates and re-optimization of the surrogate.
As a result, the calculated parameters of a well performing SBO are close to
those of a conventional optimization algorithm.

\subsection{Cost function}
\label{sec:CostFunction}

  For the optimization, we defined the discrete cost function
  \begin{align}
    \label{eqn:CostFunction}
    J (\mathbf{z}) &:= \frac{1}{2} \left\| \mathbf{z} - \mathbf{y}_d
                                   \right\|_{2}^2
                    := \frac{1}{2} \sum_{j=0}^{n_t - 1} \sum_{i=1}^{n_y}
                       \sum_{k=1}^{n_x}
                           \left( z_{jik} - y_{d, jik} \right)^2
  \end{align}
  by measuring the difference between an annual cycle $\mathbf{z} \in
  \mathbb{R}^{n_t n_y n_x}$, for example calculated during the optimization
  process, and the target annual cycle $\mathbf{y}_d \in
  \mathbb{R}^{n_t n_y n_x}$ in an Euclidean norm. Both vectors $\mathbf{z}$
  and $\mathbf{y}_d$ are indexed as $\mathbf{z} = \left( \left( \left(
  z_{jik} \right)_{k=1}^{n_x} \right)_{i=1}^{n_y} \right)_{j=0}^{n_t - 1}$
  and $\mathbf{y}_d = \left( \left( \left( y_{jik} \right)_{k=1}^{n_x}
  \right)_{i=1}^{n_y} \right)_{j=0}^{n_t - 1}$. We did not use weights in
  the Euclidean norm because we performed the following optimizations
  exclusively with synthetic data. When using real measurement data, the cost
  function \eqref{eqn:CostFunction} can be restricted by weights, for example,
  to take only discrete points in space and time for which measurements exist or
  the error variances of the measurements into account. In addition, the cost
  function can be restricted to individual tracers. \textcite{KrKhOs10}
  investigated different cost functions for biogeochemical models.

\subsection{Optimization}
\label{sec:Optimization}

  We used the SBO instead of a conventional optimization approach
  (for example gradient-based or meta-heuristics) to identify the optimal
  parameters of a biogeochemical model. These parameters are usually
  identified using a nonlinear optimization problem
  \begin{align*}
    \min_{\mathbf{u} \in U_{ad}} J \left( \mathbf{y}_f (\mathbf{u}) \right)
  \end{align*}
  with the set of admissible parameters $U_{ad} := \left\{ \mathbf{u} \in
  \mathbb{R}^{n_u}~:~\mathbf{b}_\ell \leq \mathbf{u} \leq \mathbf{b}_u
  \right\}$ and parameter boundaries $\mathbf{b}_\ell, \mathbf{b}_u \in
  \mathbb{R}^{n_u}$ with $\mathbf{b}_\ell \leq \mathbf{b}_u$. The
  inequalities with the parameter boundaries are meant component-wise. We did
  not optimize the expensive high-fidelity model ($\mathbf{y}_f$) using any
  conventional approach because the computational effort is unaffordable.

  The SBO replaces the high-fidelity model with its surrogate, whose evaluation
  is computationally cheaper than the high-fidelity model but still provides a
  reasonably accurate approximation of the high-fidelity model. There are
  several ways of constructing the surrogate, such as
  \emph{function-approximation} surrogates \parencite{QHSGVT05, SPKA01,
  SmoSch04} or \emph{physically-based} surrogates \parencite{Son03}. In the
  present paper, we used two different methods to construct the low-fidelity
  model ($\mathbf{y}_c$): a physics-based low-fidelity model and a neural
  network. These low-fidelity models are a less accurate but computationally
  more efficient approximation of the high-fidelity model. However,
  the accuracy of the low-fidelity models is not sufficient for a direct
  optimization. In order to reduce the misalignment with respect to the
  high-fidelity model, we form  the surrogate from the low-fidelity model with
  a subsequent correction. Since the surrogate approximates only the
  high-fidelity model adequately in a local environment, we also used a
  trust-region safeguard \parencite{CoGoTo00, KoBaCh10, PPKOS13} to limit the
  step size in the optimization to a certain trusted region ensuring the
  computation of a local minimum with the surrogate if both the low- and the
  high-fidelity model are sufficiently smooth. This enhances any surrogate
  optimization.

  %alg:SBO
\begin{algorithm}[t!b]
  \DontPrintSemicolon
  \KwData{$\mathbf{u}_0 \in \mathbb{R}^{n_u}$,
          $\delta_0 \in \mathbb{R}_{>0}$,
          $\delta^{\text{min}} \in \mathbb{R}_{> 0}$,
          $\gamma \in \mathbb{R}_{>0}$,
          $r_{\text{decr}} \in \mathbb{R}_{>0}$,
          $r_{\text{incr}} \in \mathbb{R}_{>0}$,
          $m_{\text{decr}} \in \mathbb{R}_{>0}$,
          $m_{\text{incr}} \in \mathbb{R}_{>0}$,
          $a_{\ell} \in \mathbb{R}_{>0}$,
          $a_{u} \in \mathbb{R}_{>0}$,
          $\delta \in \mathbb{R}_{>0}$,}
  \KwResult{$\mathbf{u}_{k_{\text{min}}} \in \mathbb{R}^{n_u}$ with $k_{\text{min}} = \min \left\{ k \in \mathbb{N}~:~\left\| \mathbf{u}_k - \mathbf{u}_{k-1} \right\|_2^2 \leq \gamma \vee \delta_k \leq \delta_k^{\text{min}} \right\}$}
  %\Begin{
    $k = 0$\;
    \While{$k = 0 \vee \left\| \mathbf{u}_k - \mathbf{u}_{k-1} \right\|_2^2 \leq \gamma \vee \delta_k \leq \delta^{\text{min}}$}{
      Evaluate high-fidelity $\mathbf{y}_f (\mathbf{u}_k)$ and low-fidelity model $\mathbf{y}_c (\mathbf{u}_k)$\;
      \tcp{Construct surrogate}
      Compute correction vector $\mathbf{a}_k \in \mathbb{R}^{n_t n_y n_x}$ with $a_{k, jil} := \begin{cases} 1.0, & \left(\mathbf{y}_f (\mathbf{u}_k) \right)_{jil} < \delta \vee \left(\mathbf{y}_c (\mathbf{u}_k) \right)_{jil} < \delta \\ a_{\ell}, & \frac{\left(\mathbf{y}_f (\mathbf{u}_k) \right)_{jil}}{\left(\mathbf{y}_c (\mathbf{u}_k) \right)_{jil}} < a_{\ell} \\ a_{u}, & \frac{\left(\mathbf{y}_f (\mathbf{u}_k) \right)_{jil}}{\left(\mathbf{y}_c (\mathbf{u}_k) \right)_{jil}} > a_{u} \\ \frac{\left(\mathbf{y}_f (\mathbf{u}_k) \right)_{jil}}{\left(\mathbf{y}_c (\mathbf{u}_k) \right)_{jil}}, & \text{else} \end{cases}$, $i \in \{1, \ldots, n_y\}, j \in \{0, \ldots, n_t - 1\}, l \in \{1, \ldots, n_x\}$\;
      $\mathbf{s}_k (\mathbf{u}) := \mathbf{a}_k \mathbf{y}_c (\mathbf{u})$\;
      \tcp{Optimize surrogate}
      $\mathbf{u}_{k+1} = \argmin_{\mathbf{u} \in U_{ab}, \left\| \mathbf{u} - \mathbf{u}_k \right\|_2 \leq \delta_k} J(\mathbf{s}_k (\mathbf{u}))$\;
      \tcp{Update trust-region radius}
      \If{$J\left(\mathbf{y}_f (\mathbf{u}_{k+1}) \right) < J\left(\mathbf{y}_f (\mathbf{u}_{k}) \right)$
      }{
        $\rho_{k} = \frac{J\left(\mathbf{y}_f (\mathbf{u}_{k+1}) \right) - J\left(\mathbf{y}_f (\mathbf{u}_{k}) \right)}{J\left(\mathbf{s}_k (\mathbf{u}_{k+1}) \right) - J\left(\mathbf{s}_k (\mathbf{u}_{k}) \right)}$\;
        $\delta_{k+1} = \begin{cases} \frac{\delta_k}{m_{\text{decr}}}, &\rho_k < r_{\text{decr}}, \\ \delta_k, & r_{\text{decr}} \leq \rho_k \leq r_{\text{incr}} \\ \delta_k \cdot m_{\text{incr}}, &\rho_k > r_{\text{incr}} \end{cases}$\;
        $k = k + 1$\;
      }
      \Else{
        $\delta_k = \frac{\delta_k}{m_{\text{decr}}}$\;
      }
    }
  %}
  \caption{Surrogate-based optimization}
  \label{alg:SBO}
\end{algorithm}

  The SBO consists of an iterative approach generating the surrogate again
  after a successful surrogate optimization. The algorithm of the SBO shown in
  Algorithm \ref{alg:SBO} can be divided into three parts. The first part
  contains the evaluation of the low- and high-fidelity model including the
  construction of the surrogate. The different low-fidelity models used in this
  paper are explained in Section \ref{sec:Low-Fidelity-Model},
  \ref{sec:Low-Fidelity-Model-ANN} as well as
  \ref{sec:Low-Fidelity-Model-ANNinit}, and the construction of the surrogate as
  a correction of the low-fidelity model is presented in Section
  \ref{sec:Low-Fidelity-Model-Correction}. In the second part of the algorithm,
  the surrogate ($\mathbf{s}_k$) is optimized using a gradient-based algorithm,
  i.e.,
  \begin{align*}
    \mathbf{u}_{k+1} &= \argmin_{\mathbf{u} \in U_{ad},
            \left\| \mathbf{u} - \mathbf{u}_k \right\| \leq \delta_k} 
        J \left( \mathbf{s}_k (\mathbf{u}) \right)
  \end{align*}
  for the iteration $k \in \mathbb{N}_0$ and trust-region radius $\delta_k \in
  \mathbb{R}_{> 0}$. If the surrogate optimization reduces the cost function
  value evaluated for the high-fidelity model, the parameter vector
  $\mathbf{u}_{k+1}$ is accepted and the trust-region radius is updated
  according to the rules below. Otherwise, the parameter vector
  $\mathbf{u}_{k+1}$ is rejected, and the surrogate optimization is performed
  again with the same surrogate but reduced trust-region radius. In analogy to
  \textcite{PPKOS13}, we used the classical rules for updating the trust-region
  radius \parencite{CoGoTo00, KoBaCh10} with slightly modified parameters,
  i.e.,
  \begin{align*}
    \delta_{k+1} &= \begin{cases} 
                      \frac{\delta_k}{m_{\text{decr}}}, & \rho_k < r_{\text{decr}} \\
                      \delta_k, & r_{\text{decr}} \leq \rho_k \leq r_{\text{incr}} \\
                      \delta_k \cdot m_{\text{incr}}, & \rho_k > r_{\text{incr}}
                \end{cases}
  \end{align*}
  with the parameters $\delta_0 = 0.06$, $r_{\text{incr}} = 0.75$,
  $r_{\text{decr}} = 0.01$ and $m_{\text{incr}}, m_{\text{decr}} \in
  \mathbb{R}_{>0}$ as well as the gain ratio
  \begin{align*}
    \rho_k &:= \frac{J \left( \mathbf{y}_f ( \mathbf{u}_{k+1}) \right) - J \left( \mathbf{y}_f ( \mathbf{u}_{k}) \right)}{J \left( \mathbf{s}_k ( \mathbf{u}_{k+1}) \right) - J \left( \mathbf{s}_k ( \mathbf{u}_{k}) \right)}.
  \end{align*}
  Consequently, the trust-region radius is decreased, on the one hand, if the
  parameter vector $\mathbf{u}_{k+1}$ is rejected, or if the improvement of the
  cost function value which was evaluated with the high-fidelity model is too
  small compared to the prediction by the surrogate and increased, on the other
  hand, if the prediction by the surrogate was adequate. The termination
  condition of the iterative approach represents the third part of the
  algorithm. We used a threshold $\gamma \in \mathbb{R}_{> 0}$ for the absolute
  step size in combination with a threshold $\delta^{\text{min}} \in
  \mathbb{R}_{> 0}$ for the trust-region radius. Alternative termination
  conditions, for example, are certain convergence criteria, an expected cost
  function value or a specific number of iterations.

  \subsubsection{Low-fidelity model using truncated spin-up}
  \label{sec:Low-Fidelity-Model}

    Our first low-fidelity model used the approach of a truncated spin-up
    (denoted as $\mathbf{y}_c^{\text{N-DOP}}$ for the N-DOP model and
    $\mathbf{y}_c^{\text{N}}$ for the N model) \parencite{PPKOS13}. In order to
    compute an approximation of the steady annual cycle using the low-fidelity
    model, we reduced the number of model years to $n_{c} = 25$  (cf. Section
    \ref{sec:ComputationSteadyAnnualCycles}). The time steps were, however,
    identical for the low- and high-fidelity model. The number of model years
    $n_{f} = 3000$ employed for the spin-up calculation of the high-fidelity
    model is much larger than $n_{c}$ for the low-fidelity model. Therefore,
    the computational effort for the low-fidelity model is significantly lower
    than for the high-fidelity model.

  \subsubsection{Low-fidelity model using prediction of an ANN}
  \label{sec:Low-Fidelity-Model-ANN}

    Our second low-fidelity model approximated the steady annual cycle with the
    prediction of an ANN \parencite{PfeSla21}. For this purpose, the ANN
    predicted the tracer concentration of a steady annual cycle approximation
    for the first time of the model year. To receive the whole trajectory as
    low-fidelity model (denoted as $\mathbf{y}_c^{\text{ANN}}$), we calculated
    the spin-up over one model year using the prediction as initial
    concentration. We used the ANN described in Section
    \ref{sec:ArtificialNeuralNetwork} including the adjustment of the prediction
    to obtain the required overall mass. Particularly, the computational effort
    to evaluate this low-fidelity model is very low because the computation of
    the trajectory for one model year is already the main effort.

  \subsubsection{Low-fidelity model using ANN as initial value generator}
  \label{sec:Low-Fidelity-Model-ANNinit}

    For our third low-fidelity model, the prediction of the ANN served as
    initial concentration for a truncated spin-up \parencite{PfeSla21}. Instead
    of using the prediction directly as low-fidelity model as in Section
    \ref{sec:Low-Fidelity-Model-ANN}, we used the ANN described in Section
    \ref{sec:ArtificialNeuralNetwork} to generate an initial concentration for
    the spin-up calculation with a reduced number of model years $n_{c} = 50$.
    The time step was the same as for the high-fidelity model. In summary, the
    low-fidelity (denoted as $\mathbf{y}_c^{\text{ANN-N}}$) model consisted of
    the ANN as initial value generator in combination with a truncated spin-up.
    This low-fidelity model reduced again the computationally effort with
    respect to the high-fidelity model.

  \subsubsection{Multiplicative response correction}
  \label{sec:Low-Fidelity-Model-Correction}

    For the three low-fidelity models, the multiplicative response correction
    reduces the misalignment of the low-fidelity model with respect to the
    high-fidelity model. \textcite{PrKoSl11, PPKOS13} have already used this
    multiplicative correction motivated by the physics-based approach of the
    low-fidelity model because the steady annual cycle calculated with the
    low-fidelity model resembles that of the high-fidelity model. For each
    iteration $k \in \mathbb{N}_0$, the multiplicative correction vector
    $\mathbf{a}_k \in \mathbb{R}^{n_t n_y n_x}$ contains the pointwise ratio of
    the high- and low-fidelity solution
    \begin{align*}
      \mathbf{a}_k &:= \frac{\mathbf{y}_f (\mathbf{u}_k)}
                            {\mathbf{y}_c (\mathbf{u}_k)}.
    \end{align*}
    In order to avoid larger entries in the correction vector, for example for
    values of the low-fidelity model close to zero or for a value of the
    low-fidelity model being several orders of magnitude smaller than the value
    of the high-fidelity model, we restrict the entries of the correction vector
    by an upper bound ($a_{u} \in \mathbb{R}_{>0}$) and a lower one ($a_{\ell}
    \in \mathbb{R}_{>0}$) and, furthermore, we set the value to $1.0$ if the
    value of the low-fidelity model or high-fidelity model is close to zero,
    i.e., falls below a threshold of $\delta = 5 \cdot 10^{-3}$. The pointwise
    multiplication of the correction vector $\mathbf{a}_k$ and the low-fidelity
    solution defines the surrogate, i.e.,
    \begin{align*}
      \mathbf{s}_k \left( \mathbf{u} \right) &:= \mathbf{a}_k 
                   \mathbf{y}_c \left( \mathbf{u} \right)
    \end{align*}
    for $\mathbf{u} \in U_{ad}$. In particular, the construction of the
    surrogate exclusively requires a single evaluation of the high-fidelity
    model. The subsequent optimization needs only the evaluation of the
    surrogate and thus of the low-fidelity model. Moreover, this construction
    of the surrogate improves its accuracy and the performance of the entire
    SBO \parencite{PPKOS13}.

  \section{Results}
  \label{sec:Results}

    We present in this section the results obtained with the three different
low-fidelity models. For the low-fidelity model described in Section
\ref{sec:Low-Fidelity-Model}, we performed a parameter optimization for both
biogeochemical models. In contrast, we used the low-fidelity models described
in Section \ref{sec:Low-Fidelity-Model-ANN} and
\ref{sec:Low-Fidelity-Model-ANNinit} only for a parameter optimization using
the N model because the neural network in Section
\ref{sec:ArtificialNeuralNetwork} was designed for the N model. In this paper,
we have used synthetic target data for the parameter optimization in order to
compare the different low-fidelity models. We assessed the optimization, on
the one hand, with the improvement of the cost function $J \left( \mathbf{y}_f
\right)$ and, on the other hand, using the accuracy of matching both the target
phosphate concentration and the optimal parameters.

\subsection{Experimental setup}
\label{sec:ExperimentalSetup}

  We configured the surrogate-based optimization uniformly for each
  low-fidelity model. Although Table \ref{table:ModelParameterValues} lists the
  parameter vectors exclusively for the N-DOP model, we used these parameter
  vectors also for the N model. For this purpose, the parameter vector $\left(
  k_w, \mu_P, K_N, K_I, \sigma_\text{DOP}, \lambda'_\text{DOP}, b \right)$ of 
  the N-DOP model had to be restricted to the parameter vector $\left( k_w,
  \mu_P, K_N, K_I, b \right)$ of the N model, i.e., the parameters
  $\sigma_\text{DOP}$ and $\lambda'_\text{DOP}$ were omitted.

  %table:ModelParameterValues
  \begin{table}[tb]
    \centering
    \caption{Parameter vectors for the biogeochemical models. Initial
             ($\mathbf{u}_0$) and optimal ($\mathbf{u}_d$) parameter vector
             used for the SBO runs, test parameter vector ($\bar{\mathbf{u}}$
             and $\tilde{\mathbf{u}}$) to assess the quality of the surrogate
             model as well as lower $(\mathbf{b}_\ell)$ and upper
             $(\mathbf{b}_u)$ bounds for the parameter vectors.}
    \label{table:ModelParameterValues}
    \begin{tabular}{c c c c c c c c}
      \hline
      Parameter & $k_w$ & $\mu_P$ & $K_N$ & $K_I$ & $\sigma_\text{DOP}$ & $\lambda'_\text{DOP}$ & $b$ \\
      \hline
      $\mathbf{u}_0$ & 0.04  &  3.5  &  0.8  &  25.0  &  0.4  &  0.3  &  0.78  \\
      $\mathbf{u}_d$ & 0.02  &  2.0  &  0.5  &  30.0  &  0.67  &  0.5  &  0.858  \\
      $\bar{\mathbf{u}}$ & 0.016 & 1.6 & 0.4 & 24.0 & 0.536 & 0.4 & 0.686 \\
      $\tilde{\mathbf{u}}$ & 0.0193 & 1.93 & 0.483 & 29.0 & 0.648 & 0.483 & 0.829 \\
      $\mathbf{b}_{\ell}$ & 0.01 & 1.0 & 0.25 & 15.0 & 0.05 & 0.25 & 0.7 \\
      $\mathbf{b}_{u}$ & 0.05 & 4.0 & 1.0 & 60.0 & 0.95 & 1.0 & 1.5 \\
      \hline
    \end{tabular}
  \end{table}

  \subsubsection{Training of the ANN}
  \label{sec:ANNTraining}

    We trained the fully connected network described in Section
    \ref{sec:ArtificialNeuralNetwork} using the parameter vectors of a Latin
    hypercube sample \parencite[cf.][]{McBeCo79}. We created the Latin
    hypercube sample with \num{1100} parameter vectors within the bounds
    ($\mathbf{b}_\ell$ and $\mathbf{b}_u$) of Table
    \ref{table:ModelParameterValues} using the routine \textsc{Lhs} of
    \textcite{PyDOE17} and computed a steady annual cycle for each parameter
    vector. For the computation of the steady annual cycle, we applied the
    marine ecosystem toolkit for optimization and simulation in 3D (Metos3D)
    \parencite{PiwSla16} and performed a spin-up over \num{10000} model years,
    starting from a global mean concentration of
    \SI{2.17}{\milli\mole\Phosphate\per\cubic\metre} for $\textrm{PO}_{4}$.
    For the training of the fully connected network, we used the parameter
    vectors of the Latin hypercube sample as input and the tracer concentration
    of the first time instant in the last, the \num{10000}th, model year as output.
    We trained the neural network over \num{1000} iterations/epochs.

  \subsubsection{Setup of the surrogate-based optimization}
  \label{sec:SetupSBO}

    For the calculation of a steady annual cycle, we applied Metos3D using
    $n_t = 45$ time steps. Except for the two low-fidelity models using the
    ANN (see Section \ref{sec:Low-Fidelity-Model-ANN} and
    \ref{sec:Low-Fidelity-Model-ANNinit}), the spin-up calculation started
    always with a constant global mean concentration of
    \SI{2.17}{\milli\mole\Phosphate\per\cubic\metre} for $\textrm{PO}_{4}$ and,
    if present, \SI{0.0001}{\milli\mole\Phosphate\per\cubic\metre} for
    $\textrm{DOP}$. For the synthetic data, we selected a random parameter
    vector $\mathbf{u}_d \in \mathbb{R}^{n_u}$ (see Table
    \ref{table:ModelParameterValues}) and computed the steady annual cycle
    $\mathbf{y}_d \in \mathbb{R}^{n_t n_y n_x}$ using the spin-up over
    \num{10000} model years with $n_t = 2880$ time steps and restricted the steady annual cycle to the corresponding time steps for $n_t = 45$, i.e.,
    \begin{align*}
      \mathbf{y}_d := \left( \mathbf{y}_i^{10000} \left( \mathbf{u}_d \right) \right)_{i \in \{0, 64, \ldots, 2816\}}.
    \end{align*}

    The SBO used the L-BFGS-B algorithm as optimization algorithm for
    the surrogate \parencite{BLNZ95, ZBLN97} implemented in
    SciPy\footnote{Scientific computing tools for Python,
    \url{https://www.scipy.org}} \parencite{VGOHRC20}. We started the SBO
    exemplary from a random parameter vector $\mathbf{u}_0 \in \mathbb{R}^{n_u}$
    shown in Table \ref{table:ModelParameterValues}. Moreover, we limited the
    admissible parameter vectors using the lower $\mathbf{b}_\ell$ and upper
    bound $\mathbf{b}_u$ (Table \ref{table:ModelParameterValues}). The SBO
    stopped if the absolute step size fell below a threshold of $\gamma = 5
    \cdot 10^{-4}$, or the trust-region radius undercut a threshold of
    $\delta^{\text{min}} = 5 \cdot 10^{-5}$. Finally, we restricted the entries
    of the multiplicative correction vector for each surrogate using the lower
    and upper bounds $a_\ell = 0.1$ and  $a_u = 5$.

\subsection{Suitability of the multiplicative correction technique}
\label{sec:SuitabilityCorrectionTechnique}

  \begin{figure}[!tb]
    \centering
    \subfloat[$\mathbf{y}_f (\bar{\mathbf{u}}) - \mathbf{y}_c^{\text{N}} (\bar{\mathbf{u}})$]{\includegraphics[width=0.315\textwidth, viewport=0 18 137 88, clip]{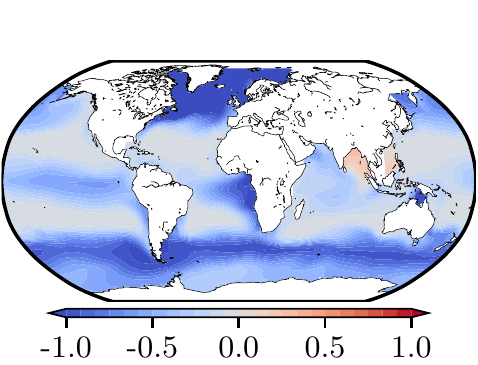}}
    \quad
    \subfloat[$\mathbf{y}_f (\bar{\mathbf{u}}) - \mathbf{y}_c^{\text{ANN}} (\bar{\mathbf{u}})$]{\includegraphics[width=0.315\textwidth, viewport=0 18 137 88, clip]{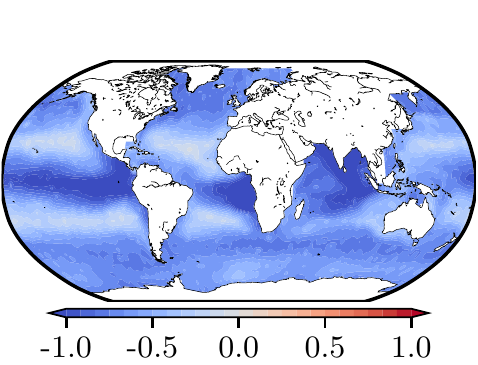}}
    \quad
    \subfloat[$\mathbf{y}_f (\bar{\mathbf{u}}) - \mathbf{y}_c^{\text{ANN-N}} (\bar{\mathbf{u}})$]{\includegraphics[width=0.315\textwidth, viewport=0 18 137 88, clip]{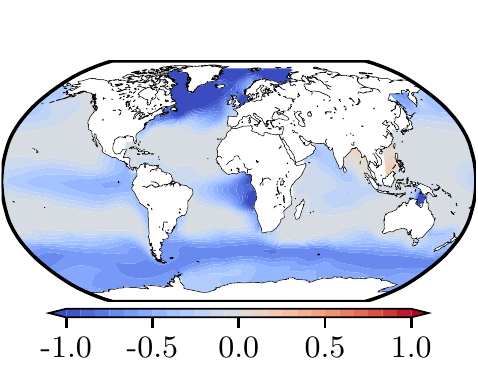}}
    \quad
    \subfloat[$\mathbf{y}_f (\bar{\mathbf{u}}) - \mathbf{s}_0^{\text{N}} (\bar{\mathbf{u}})$]{\includegraphics[width=0.315\textwidth, viewport=0 18 137 88, clip]{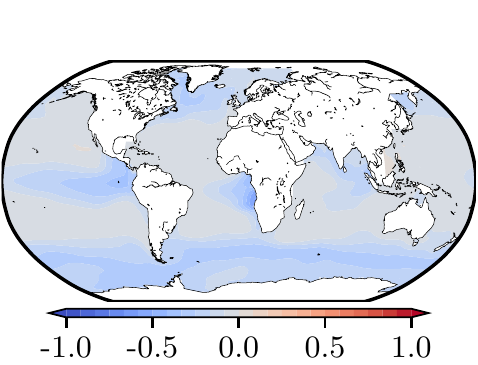}}
    \quad
    \subfloat[$\mathbf{y}_f (\bar{\mathbf{u}}) - \mathbf{s}_0^{\text{ANN}} (\bar{\mathbf{u}})$]{\includegraphics[width=0.315\textwidth, viewport=0 18 137 88, clip]{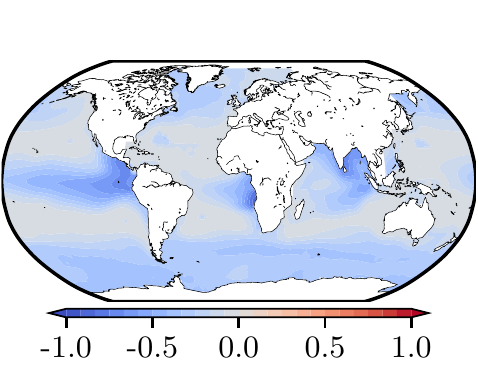}}
    \quad
    \subfloat[$\mathbf{y}_f (\bar{\mathbf{u}}) - \mathbf{s}_0^{\text{ANN-N}} (\bar{\mathbf{u}})$]{\includegraphics[width=0.315\textwidth, viewport=0 18 137 88, clip]{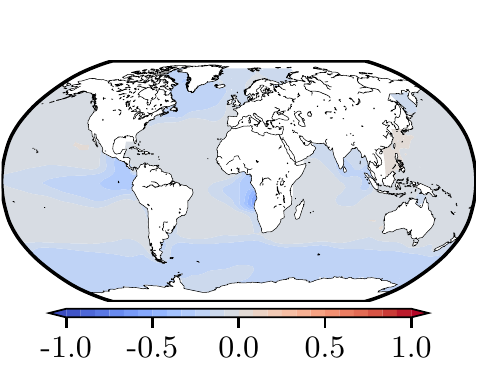}} \\
    \subfloat[$\mathbf{y}_f (\tilde{\mathbf{u}}) - \mathbf{y}_c^{\text{N}} (\tilde{\mathbf{u}})$]{\includegraphics[width=0.315\textwidth, viewport=0 18 137 88, clip]{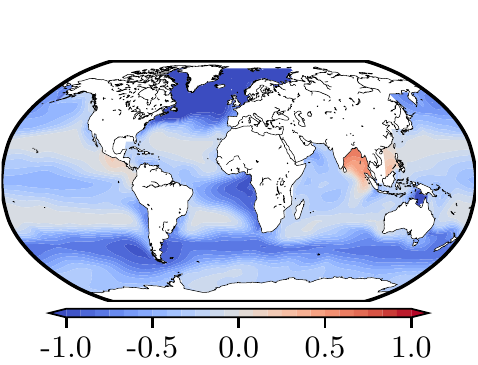}}
    \quad
    \subfloat[$\mathbf{y}_f (\tilde{\mathbf{u}}) - \mathbf{y}_c^{\text{ANN}} (\tilde{\mathbf{u}})$]{\includegraphics[width=0.315\textwidth, viewport=0 18 137 88, clip]{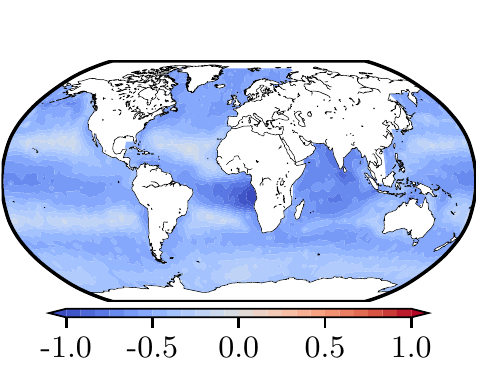}}
    \quad
    \subfloat[$\mathbf{y}_f (\tilde{\mathbf{u}}) - \mathbf{y}_c^{\text{ANN-N}} (\tilde{\mathbf{u}})$]{\includegraphics[width=0.315\textwidth, viewport=0 18 137 88, clip]{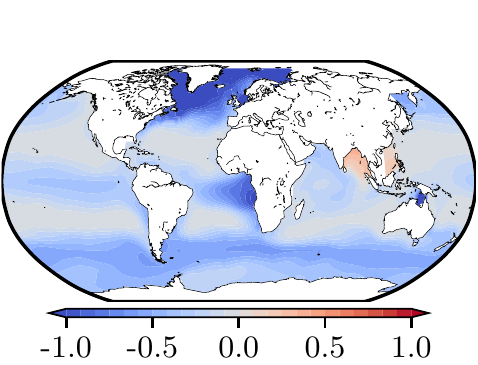}} \\
    \subfloat[$\mathbf{y}_f (\tilde{\mathbf{u}}) - \mathbf{s}_0^{\text{N}} (\tilde{\mathbf{u}})$]{\includegraphics[width=0.315\textwidth, viewport=0 18 137 88, clip, valign=t]{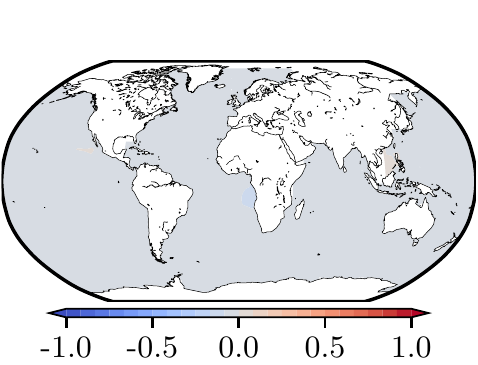}
    \vphantom{\includegraphics[width=0.315\textwidth, viewport=0 2 137 88, clip, valign=t]{Figures/SBO_N_Surrogate_2.pdf}}}
    \;
    \subfloat[$\mathbf{y}_f (\tilde{\mathbf{u}}) - \mathbf{s}_0^{\text{ANN}} (\tilde{\mathbf{u}})$]{\includegraphics[width=0.315\textwidth, viewport=0 2 137 88, clip, valign=t]{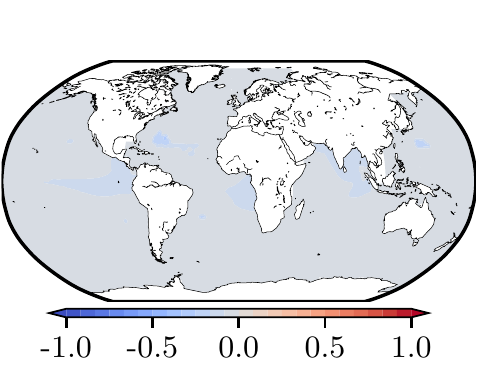}}
    \;
    \subfloat[$\mathbf{y}_f (\tilde{\mathbf{u}}) - \mathbf{s}_0^{\text{ANN-N}} (\tilde{\mathbf{u}})$]{\vphantom{\includegraphics[width=0.315\textwidth, viewport=0 2 137 88, clip, valign=t]{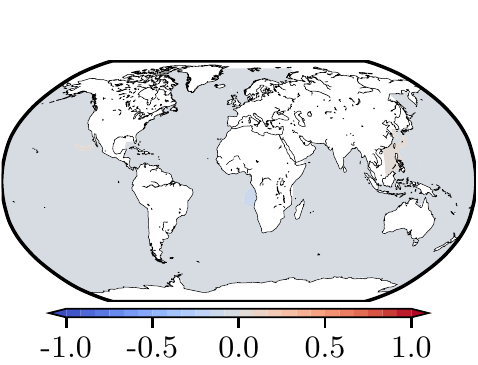}}
    \includegraphics[width=0.315\textwidth, viewport=0 18 137 88, clip, valign=t]{Figures/SBO_N-ANN_Surrogate_2.pdf}}
    \caption{Difference between the high-fidelity model $\mathbf{y}_f$ and the
             low-fidelity model $\mathbf{y}_c$ as well as the surrogate
             $\mathbf{s}_0$ for the three different low-fidelity models
             (truncated-spin-up $\mathbf{y}_c^{\text{N}}$ (left, Section
             \ref{sec:Low-Fidelity-Model}), ANN prediction
             $\mathbf{y}_c^{\text{ANN}}$ (middle, Section
             \ref{sec:Low-Fidelity-Model-ANN}) and ANN prediction as
             initial concentration for the truncated spin-up
             $\mathbf{y}_c^{\text{ANN-N}}$ (right, Section
             \ref{sec:Low-Fidelity-Model-ANNinit}) as low-fidelity
             model). For the two parameter vectors $\bar{\mathbf{u}}$ and
             $\tilde{\mathbf{u}}$, the phosphate concentration difference are
             shown on the surface (\SIrange{0}{50}{\metre}) at the first time
             step of the model year (in January). The surrogate was built
             using the parameter vector $\mathbf{u}_d$, respectively.}
    \label{fig:LowFidelityModels}
  \end{figure}

  The multiplicative response correction was suitable for the three
  low-fidelity models. Figure \ref{fig:LowFidelityModels} indicates the
  improvement of the low-fidelity model accuracy using the multiplicative
  response correction. Although the concentrations of the low-fidelity models
  differed significantly from each other in some areas, the approximations of
  the high-fidelity model were very similar after the multiplicative
  correction. Particularly, the surrogate approximated the high-fidelity model
  better for each low-fidelity model if the parameter vector was closer to the
  parameter vector used to construct the surrogate. The approximation was,
  nevertheless, slightly worse using the prediction of the ANN as low-fidelity
  model while the other two low-fidelity models showed only marginal
  differences in the approximations after the correction. Except for the
  low-fidelity model using the ANN prediction as detailed in Section
  \ref{sec:Results-ANN}, the surrogates provided an appropriate approximation
  of the high-fidelity model in the neighborhood of the parameter vector used
  to construct the surrogate.

\subsection{Low-fidelity model as truncated spin-up}
\label{sec:Results_TruncatedSpin-up}

  In this section, we present the parameter identification with the SBO using
  the low-fidelity model based on the truncated spin-up (see Section
  \ref{sec:Low-Fidelity-Model}) for both biogeochemical models. For the update
  of the trust-region radius, we applied the parameter $m_{\text{incr}} = 3$ for
  both biogeochemical models, as well as the parameter $m_{\text{decr}} = 20$
  for the N-DOP model and $m_{\text{decr}} = 3$ for the N model.

  \subsubsection{N-DOP model}
  \label{sec:Results-N-DOP}

    \begin{figure}[tb]
      \centering
      \subfloat[Cost function value $J\left( \mathbf{y}_f \right)$.\label{fig:N-DOP_ConvergenceCostfunction}]{\includegraphics{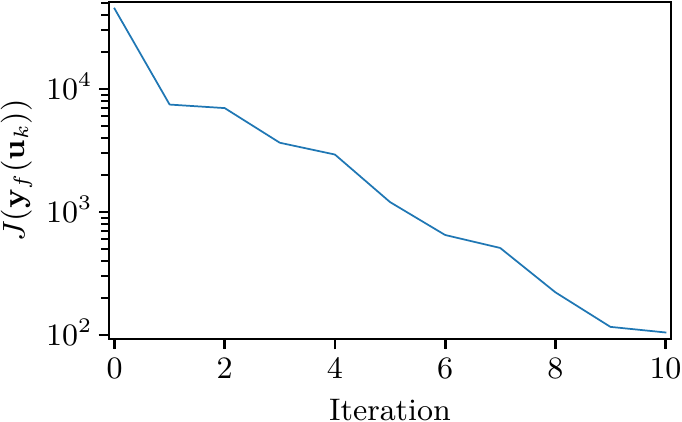}}
      \quad
      \subfloat[Step-size norm of the trust-region radius $\delta_k$.]{\includegraphics{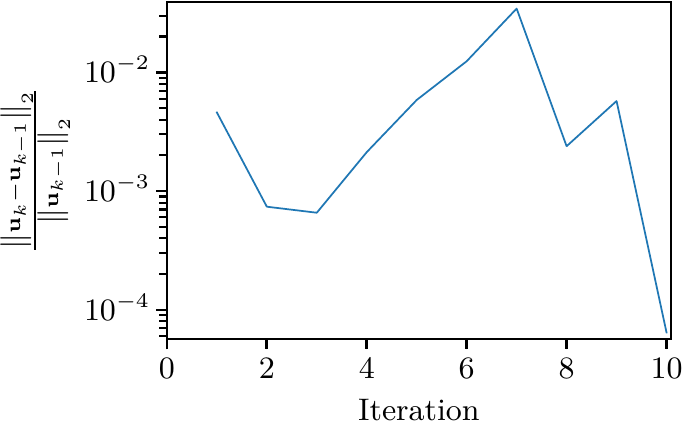}}
      \caption{Convergence of the cost function value and the step-size norm of
               the trust-region radius for the exemplary SBO run with the
               truncated spin-up as low-fidelity model
               ($\mathbf{y}_c^{\text{N-DOP}}$) and the N-DOP model.}
      \label{fig:N-DOP_Convergence}
    \end{figure}

    \begin{figure}[tb]
      \centering
      \subfloat{\includegraphics[width=0.315\textwidth]{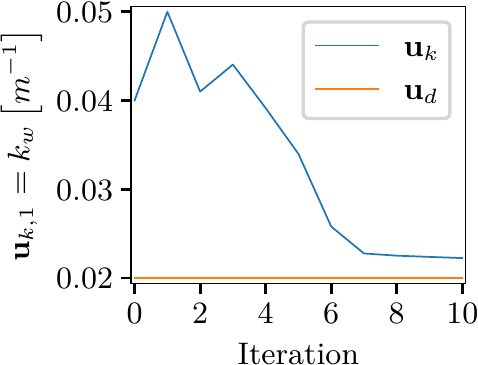}}
      \quad
      \subfloat{\includegraphics[width=0.315\textwidth]{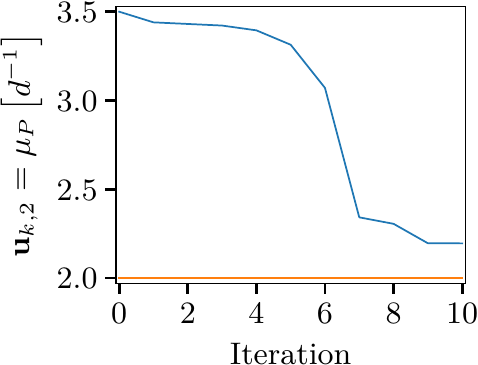}}
      \quad
      \subfloat{\includegraphics[width=0.315\textwidth]{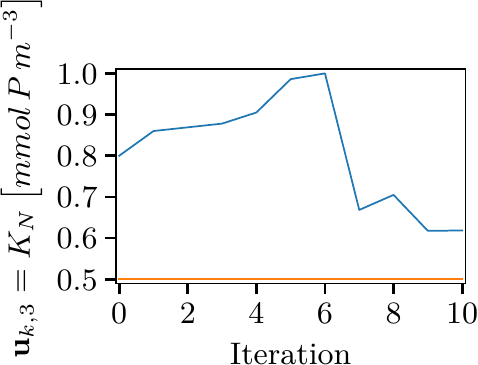}}
      \quad
      \subfloat{\includegraphics[width=0.315\textwidth]{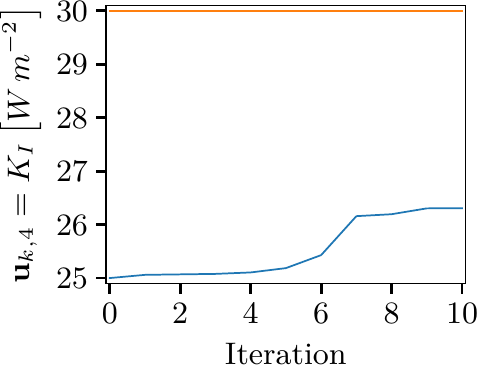}}
      \quad
      \subfloat{\includegraphics[width=0.315\textwidth]{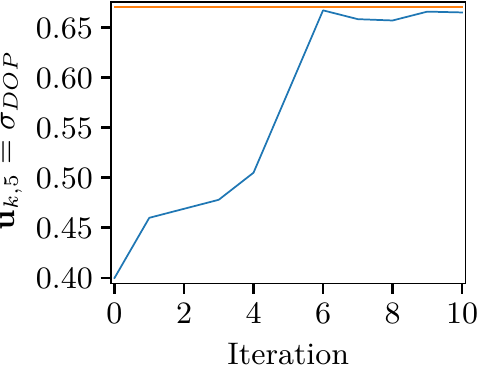}}
      \quad
      \subfloat{\includegraphics[width=0.315\textwidth]{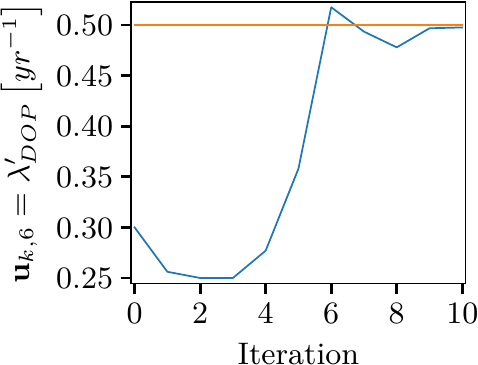}}
      \quad
      \subfloat{\includegraphics[width=0.315\textwidth]{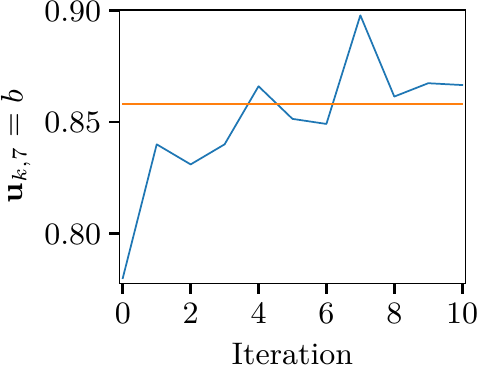}}
      \caption{Convergence of the single parameter values $\mathbf{u}_{k,i}$
               for each iteration of the exemplary SBO run with the truncated
               spin-up as low-fidelity model ($\mathbf{y}_c^{\text{N-DOP}}$)
               and the N-DOP model.}
      \label{fig:N-DOPConvergenceParameter}
    \end{figure}

\begin{figure}[p]
      \centering
      \subfloat[$\mathbf{y}_d$]{\includegraphics[width=0.315\textwidth, viewport=0 2 137 88, clip]{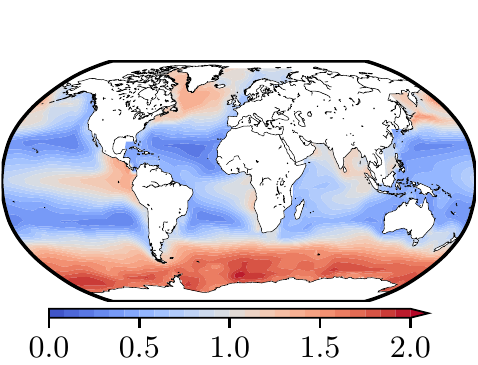}}
      \quad
      \subfloat[$\mathbf{y}_f (\mathbf{u}_0)$]{\includegraphics[width=0.315\textwidth, viewport=0 2 137 88, clip]{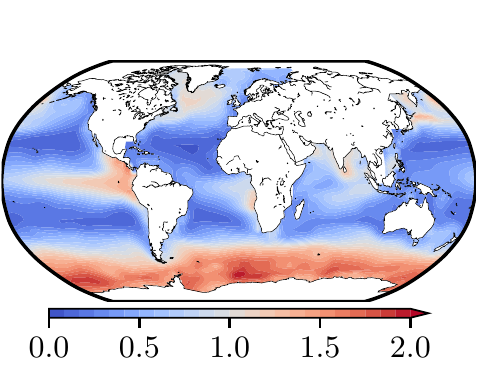}}
      \quad
      \subfloat[$\mathbf{y}_f (\mathbf{u}_2)$]{\includegraphics[width=0.315\textwidth, viewport=0 2 137 88, clip]{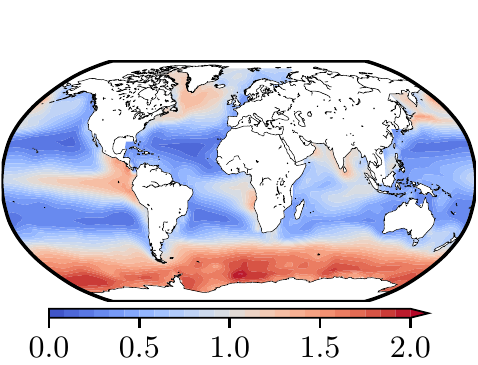}}
      \quad
      \subfloat[$\mathbf{y}_f (\mathbf{u}_5)$]{\includegraphics[width=0.315\textwidth, viewport=0 2 137 88, clip]{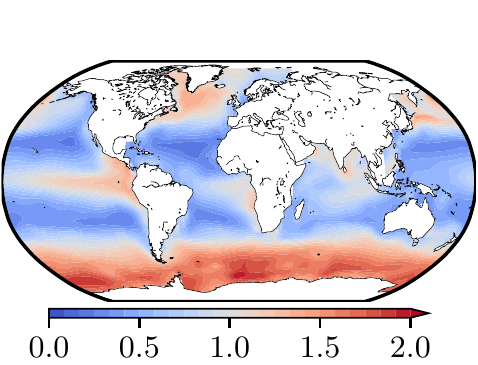}}
      \quad
      \subfloat[$\mathbf{y}_f (\mathbf{u}_8)$]{\includegraphics[width=0.315\textwidth, viewport=0 2 137 88, clip]{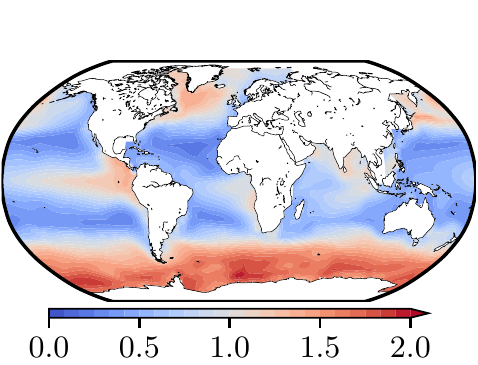}}
      \quad
      \subfloat[$\mathbf{y}_f (\mathbf{u}_{10})$]{\includegraphics[width=0.315\textwidth, viewport=0 2 137 88, clip]{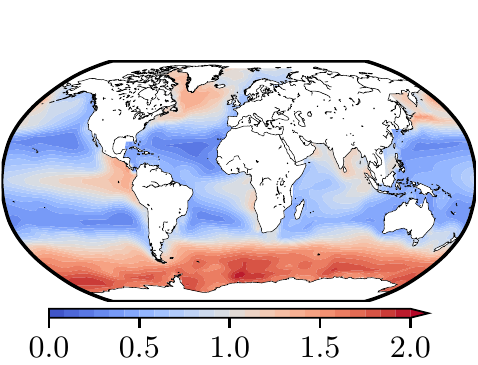}}
      \caption{High-fidelity model output $\mathbf{y}_f$ obtained by the
               exemplary SBO run with the truncated spin-up as low-fidelity
               model ($\mathbf{y}_c^{\text{N-DOP}}$) and the N-DOP model at the
               beginning and after two, five, eight and ten iterations (i.e.,
               evaluated with the parameter vectors $\mathbf{u}_0$,
               $\mathbf{u}_2$, $\mathbf{u}_5$, $\mathbf{u}_8$ and
               $\mathbf{u}_{10}$) as well as the target data $\mathbf{y}_d$.
               Shown are the phosphate concentrations on the surface layer
               (\SIrange{0}{50}{\metre}) at the first time step of the model
               year (in January).}
      \label{fig:N-DOP_Surface}
    \end{figure}

\begin{figure}[p]
      \centering
      \subfloat[Location \ang{120.9375} W, \ang{30.9375} N]{\includegraphics{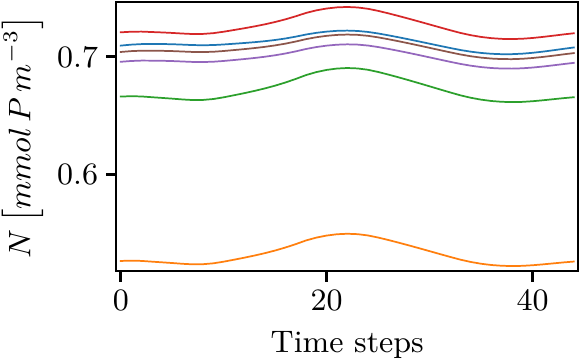}}
      \quad
      \subfloat[Location \ang{90.0} E, \ang{0.0} N]{\includegraphics{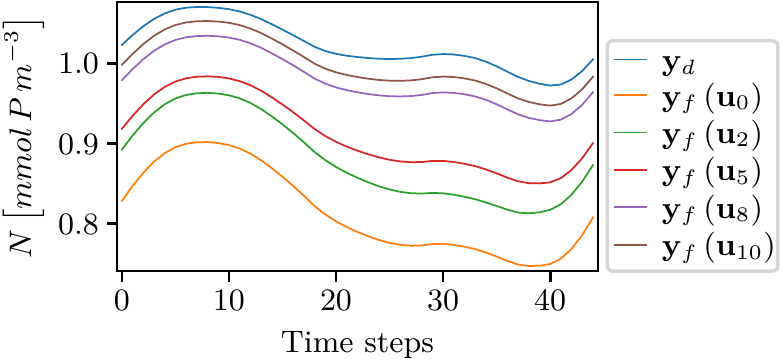}}
      \caption{Annual cycles of phosphate of the high-fidelity model output
               $\mathbf{y}_f$ obtained by the exemplary SBO run with the
               truncated spin-up as low-fidelity model
               ($\mathbf{y}_c^{\text{N-DOP}}$) and the N-DOP model at
               the beginning and after two, five, eight and ten iterations
               (i.e., evaluated with the parameter vectors $\mathbf{u}_0$,
               $\mathbf{u}_2$, $\mathbf{u}_5$, $\mathbf{u}_8$ and
               $\mathbf{u}_{10}$) as well as the target data $\mathbf{y}_d$ on
               the surface layer (\SIrange{0}{50}{\metre}) for two distinct
               locations.}
      \label{fig:N-DOPAnnualCycle}
    \end{figure}

    The parameter identification using the SBO identified six out of seven
    model parameters of the N-DOP model. The steady annual cycles calculated
    for each iteration of the SBO converged to the target
    concentration $\mathbf{y}_d$ (Figure \ref{fig:N-DOP_Convergence}).
    Furthermore, the single parameter values converged against the values of
    the optimal parameter vector $\mathbf{u}_d$ (see Figure
    \ref{fig:N-DOPConvergenceParameter}), except for parameter $K_I$.
    Especially in the iterations four to eight, the SBO reduced the difference
    to the optimal parameter vector whereas even the step size increased
    significantly. The SBO terminated after ten iterations because the step
    size fell below the threshold of $\gamma = 5 \cdot 10^{-5}$ (i.e., $\left\|
    \mathbf{u}_{10} - \mathbf{u}_9 \right\|_2^2 < \gamma$). Although the main
    patterns of the target phosphate concentration are already visible for
    the solution $\mathbf{y}_f (\mathbf{u}_0)$ of the initial parameter vector
    $\mathbf{u}_0$, Figures \ref{fig:N-DOP_Surface} and
    \ref{fig:N-DOPAnnualCycle} show the further improvement of the match (for
    instance in the North Atlantic on the surface layer) during the
    optimization.

    The SBO was computationally very efficient. The exemplary SBO run required
    only \num{15} high-fidelity model evaluations, whereby the surrogate
    optimization of four iterations were discarded due to a larger cost
    function value after the optimization. In contrast, the surrogate
    optimization needed altogether more than \num{6000} evaluations of the
    surrogate. This large number of surrogate evaluations resulted from the
    limitation of the iterations of the surrogate optimization to \num{100}.
    Limiting the number of iterations to \num{10}, the SBO run converged
    with a cost function value of about \num{650} (instead of about \num{100}
    using a limit of \num{100} iterations as illustrated by Figure
    \ref{fig:N-DOP_ConvergenceCostfunction}) but with a significantly lower number of about
    \num{1500} surrogate evaluations.

  \subsubsection{N model}
  \label{sec:Results-N}

    \begin{figure}[!t]
      \centering
      \subfloat[Cost function value $J\left( \mathbf{y}_f \right)$.\label{fig:N_ConvergenceCostfunction}]{\includegraphics{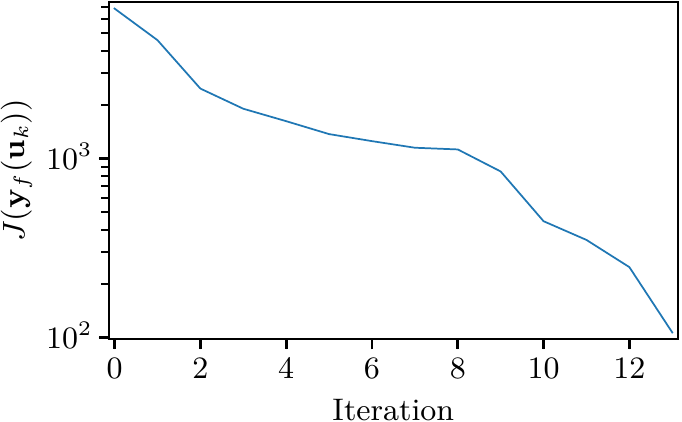}}
      \quad
      \subfloat[Step-size norm of the trust-region radius $\delta_k$.\label{fig:N_ConvergenceStepSizeNorm}]{\includegraphics{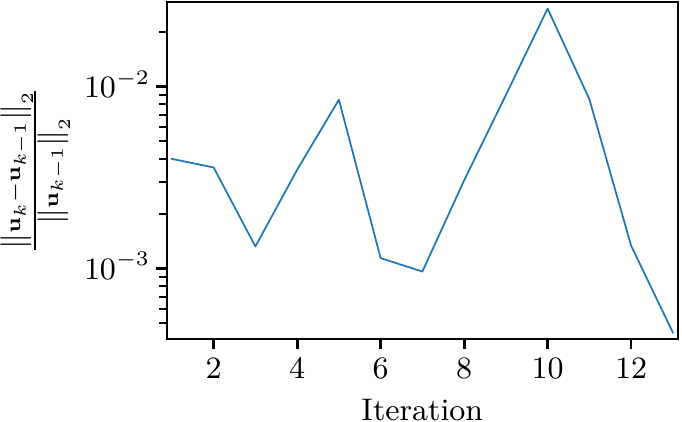}}
      \caption{Convergence of the cost function value and the step-size norm
               of the trust-region radius for the exemplary SBO run with the
               truncated spin-up as low-fidelity model
               ($\mathbf{y}_c^{\text{N}}$) and the N model.}
      \label{fig:N_Convergence}
    \end{figure}

    \begin{figure}[!t]
      \centering
      \subfloat{\includegraphics[width=0.315\textwidth]{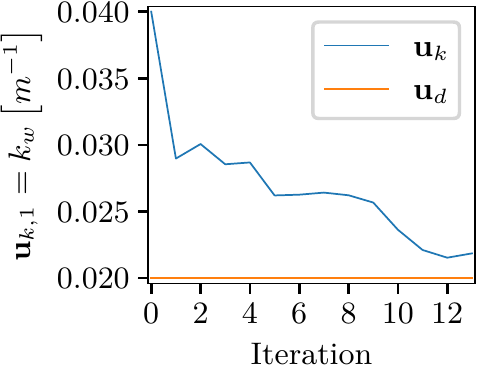}}
      \quad
      \subfloat{\includegraphics[width=0.315\textwidth]{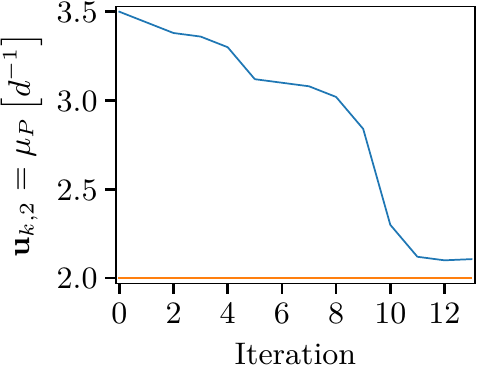}}
      \quad
      \subfloat{\includegraphics[width=0.315\textwidth]{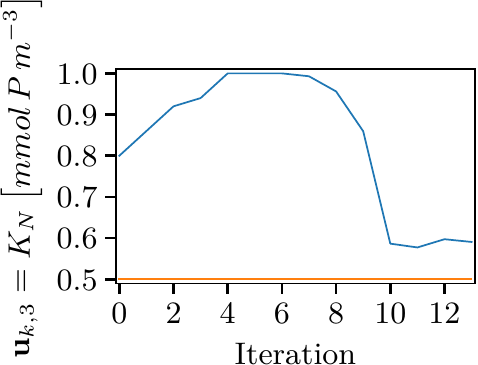}}
      \quad
      \subfloat{\includegraphics[width=0.315\textwidth]{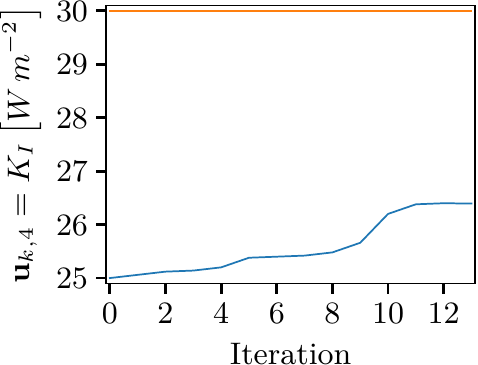}}
      \quad
      \subfloat{\includegraphics[width=0.315\textwidth]{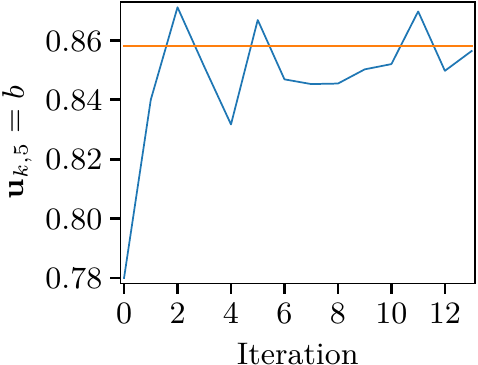}}
      \caption{Convergence of the single parameter values $\mathbf{u}_{k,i}$
               for each iteration of the exemplary SBO run with the truncated
               spin-up as low-fidelity model ($\mathbf{y}_c^{\text{N}}$) and
               the N model.}
      \label{fig:N_ConvergenceParameter}
    \end{figure}

    \begin{figure}[p]
      \centering
      \subfloat[$\mathbf{y}_d$]{\includegraphics[width=0.315\textwidth, viewport=0 2 137 88, clip]{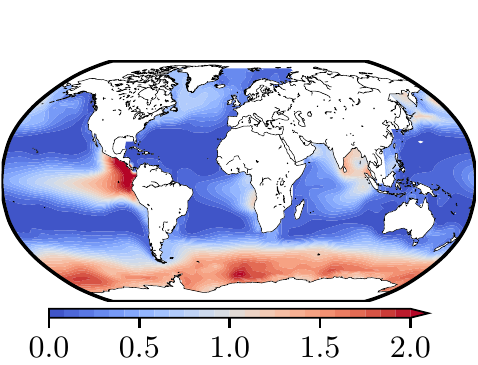}}
      \quad
      \subfloat[$\mathbf{y}_f (\mathbf{u}_0)$]{\includegraphics[width=0.315\textwidth, viewport=0 2 137 88, clip]{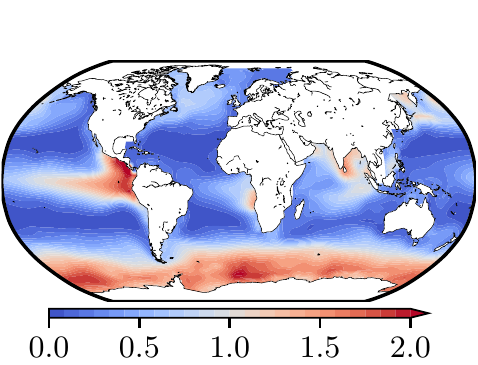}}
      \quad
      \subfloat[$\mathbf{y}_f (\mathbf{u}_4)$]{\includegraphics[width=0.315\textwidth, viewport=0 2 137 88, clip]{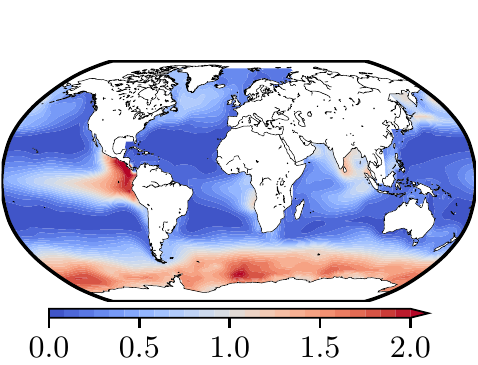}}
      \quad
      \subfloat[$\mathbf{y}_f (\mathbf{u}_7)$]{\includegraphics[width=0.315\textwidth, viewport=0 2 137 88, clip]{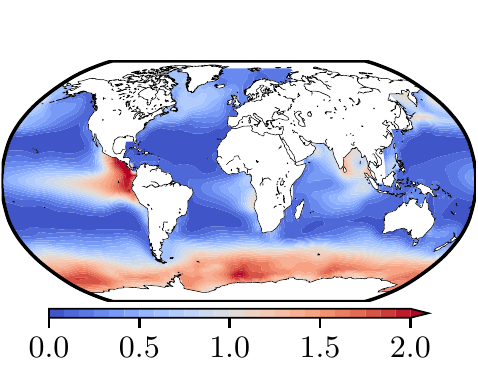}}
      \quad
      \subfloat[$\mathbf{y}_f (\mathbf{u}_{10})$]{\includegraphics[width=0.315\textwidth, viewport=0 2 137 88, clip]{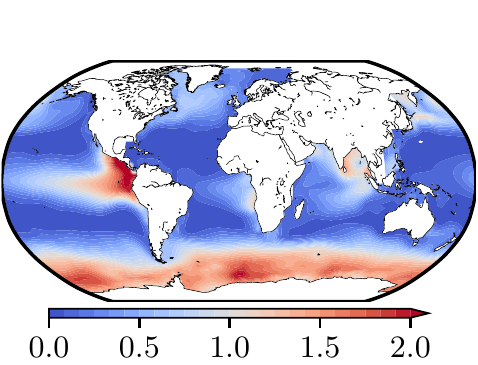}}
      \quad
      \subfloat[$\mathbf{y}_f (\mathbf{u}_{13})$]{\includegraphics[width=0.315\textwidth, viewport=0 2 137 88, clip]{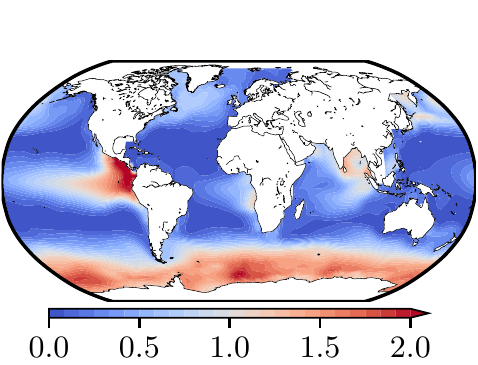}}
      \caption{High-fidelity model output $\mathbf{y}_f$ obtained by the
               exemplary SBO run with the truncated spin-up as low-fidelity
               model ($\mathbf{y}_c^{\text{N}}$) and the N model at the
               beginning and after \num{4}, \num{7}, \num{10} and \num{13}
               iterations (i.e., evaluated with the parameter vectors
               $\mathbf{u}_0$, $\mathbf{u}_4$, $\mathbf{u}_7$,
               $\mathbf{u}_{10}$ and $\mathbf{u}_{13}$) as well as the target
               data $\mathbf{y}_d$. Shown are the phosphate concentrations on
               the surface layer (\SIrange{0}{50}{\metre}) at the first time
               step of the model year (in January).}
      \label{fig:N_Surface}
    \end{figure}

    \begin{figure}[p]
      \centering
      \subfloat[Location \ang{120.9375} W, \ang{30.9375} N]{\includegraphics{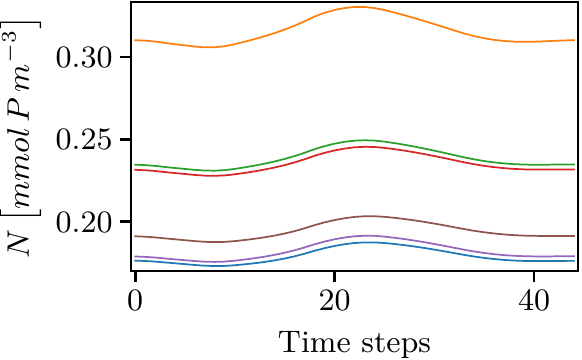}}
      \quad
      \subfloat[Location \ang{90.0} E, \ang{0.0} N]{\includegraphics{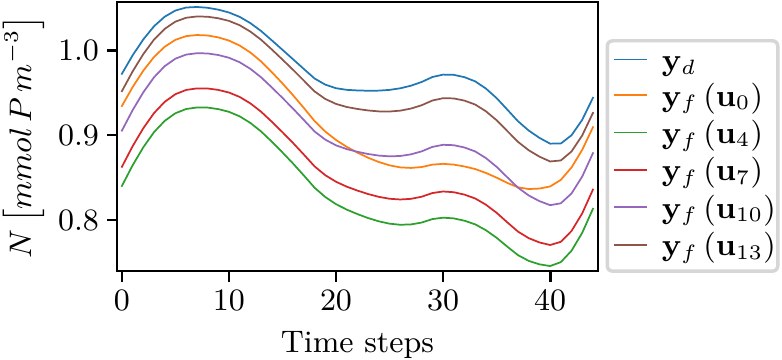}}
      \caption{Annual cycles of phosphate of the high-fidelity model output
               $\mathbf{y}_f$ obtained by the exemplary SBO run with the
               truncated spin-up as low-fidelity model
               ($\mathbf{y}_c^{\text{N}}$) and the N model at the
               beginning and after \num{4}, \num{7}, \num{10} and \num{13}
               iterations (i.e., evaluated with the parameter vectors
               $\mathbf{u}_0$, $\mathbf{u}_4$, $\mathbf{u}_7$,
               $\mathbf{u}_{10}$ and $\mathbf{u}_{13}$) as well as the target
               data $\mathbf{y}_d$ on the surface layer
               (\SIrange{0}{50}{\metre}) for two distinct locations.}
      \label{fig:N_AnnualCycle}
    \end{figure}

    The parameters converged to the optimal parameter $\mathbf{u}_d$ using the
    SBO for the N model with the exception of parameter $K_I$. The cost
    function values $J(\mathbf{y}_f)$ in Figure \ref{fig:N_ConvergenceCostfunction}
    indicate a solution close to the target data $\mathbf{y}_d$. Particularly,
    the SBO identified four out of five model parameters as detailed in Figure
    \ref{fig:N_ConvergenceParameter}. Although the first iteration already
    determined the parameter $b$, the SBO required twelve iterations to
    identify the other three parameters. Thereby, the algorithm repeatedly
    increased the step size (Figure \ref{fig:N_ConvergenceStepSizeNorm}). As a
    result of falling below the threshold $\delta^{\text{min}}$ for the
    trust-region radius, the SBO terminated after \num{13} iterations.
    According to Figures \ref{fig:N_Surface} and \ref{fig:N_AnnualCycle}, the
    main patterns of the target phosphate concentration are apparent for the
    tracer concentrations of the selected parameter vectors. However, the SBO
    improved the match sequentially (for example in the Indian Ocean, see also
    Figure \ref{fig:N_ConvergenceCostfunction}). Nevertheless, local worsening
    occurred occasionally compared to the previously calculated concentration
    (cf. Figure \ref{fig:N_AnnualCycle}).

    The computational costs of the SBO were low. The exemplary SBO run required
    only \num{34} high-fidelity model evaluations and about \num{6000}
    surrogate evaluations. Compared to the small number of \num{13} iterations,
    more than half of the high-fidelity model evaluations resulted from the
    calculation of the reference cost function value for parameter vectors that
    had been calculated by the surrogate optimization but did not cause a
    reduction of the cost function. Analogous to the SBO using the N-DOP model
    in the previous section \ref{sec:Results-N-DOP}, the number of surrogate
    evaluations could be reduced by limiting the optimization iterations of the
    surrogate optimization to \num{10}. As a consequence, the final cost
    function value increased significantly to over \num{1000} for such an
    optimization run compared to a value of about \num{100} when a limit of
    \num{100} iterations of the surrogate optimization was used as shown in
    Figure \ref{fig:N_ConvergenceCostfunction}.

\subsection{Prediction of an ANN as low-fidelity model}
\label{sec:Results-ANN}

  \begin{figure}[!t]
    \centering
    \subfloat[Location \ang{120.9375} W, \ang{30.9375} N]{\includegraphics{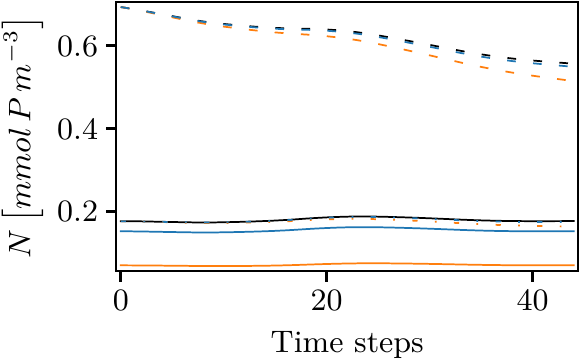}}
    \quad
    \subfloat[Location \ang{90.0} E, \ang{0.0} N]{\includegraphics{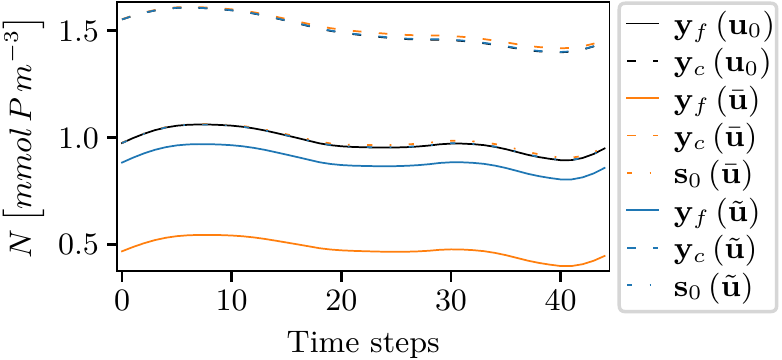}}
    \caption{Annual cycle of phosphate for the high- ($\mathbf{y}_f$),
             low-fidelity model ($\mathbf{y}_c^{\text{ANN}}$) and surrogate
             ($\mathbf{s}_0$) on the surface layer (\SIrange{0}{50}{\metre})
             for two distinct locations. Shown are the annual cycles for three
             parameter values $\mathbf{u}_0$ (as ``reference point''),
             $\bar{\mathbf{u}}$ (a neighbouring point) and $\tilde{\mathbf{u}}$
             (a point in a closer vicinity). The surrogate was built using
             parameter vector $\mathbf{u}_0$, the reason why the solution of the surrogate is
             omitted at the reference point.}
    \label{fig:N_ANN_AnnualCycle}
  \end{figure}

  Using the prediction of the ANN defined in Section
  \ref{sec:ArtificialNeuralNetwork} as low-fidelity model (see Section
  \ref{sec:Low-Fidelity-Model-ANN}) was unsuitable to construct a reliable
  surrogate in conjunction with the multiplicative response correction for the
  SBO with the N model. Figure \ref{fig:N_ANN_AnnualCycle} demonstrates the
  nearly identical phosphate concentrations predicted by the ANN for the three
  parameter vectors $\mathbf{u}_0, \bar{\mathbf{u}}, \tilde{\mathbf{u}} \in
  \mathbb{R}^{n_u}$ defined in Table \ref{table:ModelParameterValues}.
  Consequently, the steady annual cycles calculated as spin-up over one model
  year using the prediction as an initial concentration differed only marginally
  from each other due to the different model parameters. Instead of the
  high-fidelity solution of the corresponding parameter vector, the surrogate,
  therefore, in each case, approximated the high-fidelity solution for
  parameter vector $\mathbf{u}_0$ with which surrogate $\mathbf{s}_0$ was
  constructed. We observed a similar behavior using random parameter vectors
  in a small area around the reference parameters used to construct the
  surrogate.

  \begin{figure}[tb]
    \centering
    \includegraphics{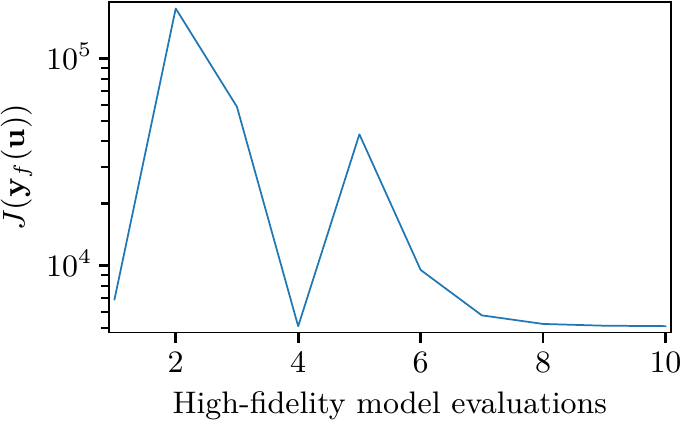}
    \caption{Convergence of the cost function value $J\left( \mathbf{y}_f
             \right)$ for the exemplary SBO run with the N model using the
             prediction of the ANN as low-fidelity model
             ($\mathbf{y}_c^{\text{ANN}}$). The iterations \num{0} and \num{1}
             (with parameter vectors $\mathbf{u}_0$ and $\mathbf{u}_1$) of the
             SBO run correspond to \num{1} and \num{4} evaluations of the
             high-fidelity model, respectively.}
    \label{fig:N_ANN_Convergence}
  \end{figure}

  \begin{figure}[tb]
    \centering
    \subfloat[$\mathbf{y}_d$]{\includegraphics[width=0.315\textwidth, viewport=0 2 137 88, clip]{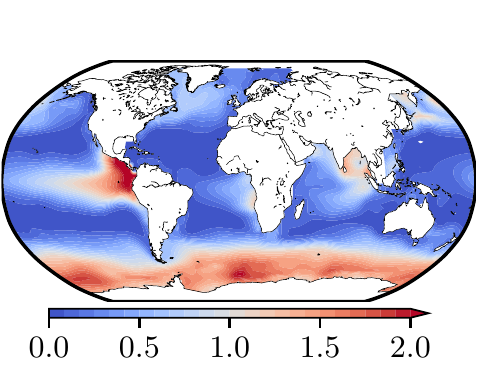}}
    \quad
    \subfloat[$\mathbf{y}_f (\mathbf{u}_0)$]{\includegraphics[width=0.315\textwidth, viewport=0 2 137 88, clip]{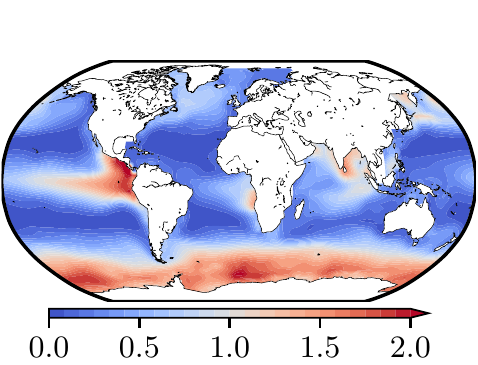}}
    \quad
    \subfloat[$\mathbf{y}_f (\mathbf{u}_1)$]{\includegraphics[width=0.315\textwidth, viewport=0 2 137 88, clip]{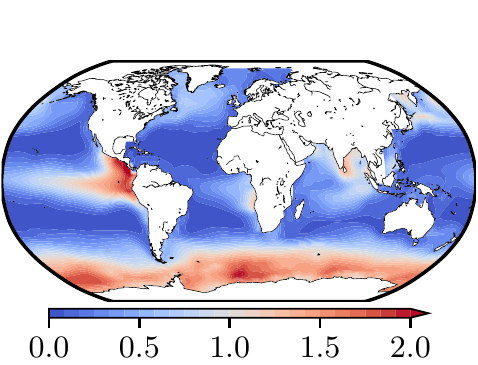}}
    \caption{High-fidelity model output $\mathbf{y}_f$ obtained by the
             exemplary SBO run with the N model using the prediction of the ANN
             as low-fidelity model ($\mathbf{y}_c^{\text{ANN}}$) at the
             beginning and after one iteration (i.e., evaluated with the
             parameter vectors $\mathbf{u}_0$ and $\mathbf{u}_{1}$) as well as
             the target data $\mathbf{y}_d$. Shown are the phosphate
             concentrations on the surface layer (\SIrange{0}{50}{\metre}) at
             the first time step of the model year (in January).}
    \label{fig:N_ANN_Surface}
  \end{figure}

  The SBO did not identify any model parameters. After three surrogate
  optimizations, the SBO calculated with surrogate $\mathbf{s}_0$ just one
  parameter vector $\mathbf{u}_1 \in \mathbb{R}^{n_u}$ for which the
  corresponding reference cost function value was smaller than for the initial
  parameter vector $\mathbf{u}_0$ (Figure \ref{fig:N_ANN_Convergence}). The
  subsequent surrogate optimizations with surrogate $\mathbf{s}_1$ did not yield
  a parameter vector whose high-fidelity model solution approximated the target
  data better. Although the main pattern of the target phosphate concentration
  for the high-fidelity solution corresponding to the parameter vector
  $\mathbf{u}_0$ as well as $\mathbf{u}_1$ were evident as seen in Figure
  \ref{fig:N_ANN_Surface}, there were many differences, such as in the Pacific,
  and, consequently, parameter vector $\mathbf{u}_1$ did not provide a
  satisfactory identification of the optimal parameter vector $\mathbf{u}_d$.
  After only one iteration and ten high-fidelity model evaluations, the SBO
  terminated because the trust-region radius was decreased in every step using
  $m_{\text{decr}} = 4$ and, hence, fell below the threshold
  $\delta^{\text{min}}$ after ten steps.

\subsection{ANN as initial value generator for the low-fidelity model}
\label{sec:Results-ANNInitialValue}

  \begin{figure}[!tb]
    \centering
    \subfloat[Generation of initial values with an ANN for a spin-up over \num{50} model years as low-fidelity model ($\mathbf{y}_c^{\text{ANN-N}}$).]{\includegraphics{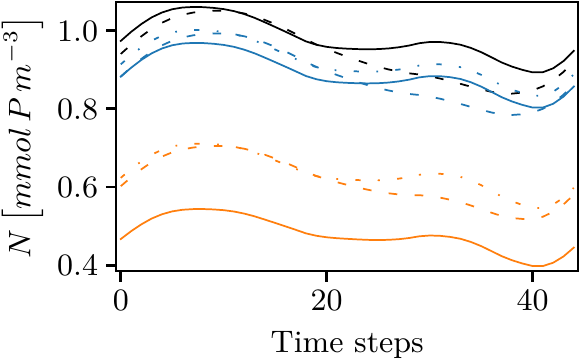}}
    \quad
    \subfloat[Spin-up over \num{25} model years using constant initial values as low-fidelity model ($\mathbf{y}_c^{\text{N}}$).]{\includegraphics{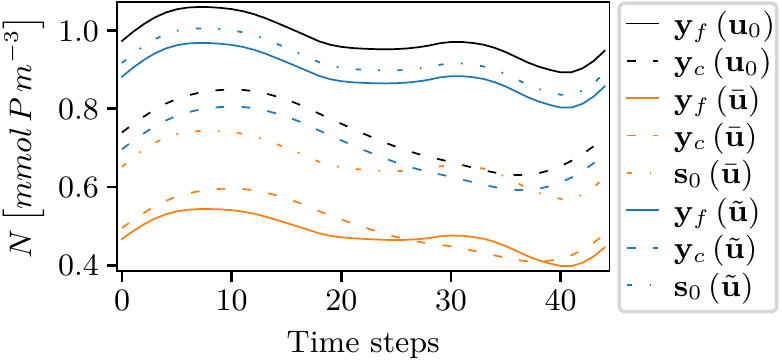}}
    \caption{Annual cycle of phosphate for the high- ($\mathbf{y}_f$),
             low-fidelity model ($\mathbf{y}_c$) and surrogate ($\mathbf{s}_0$)
             on the surface layer (\SIrange{0}{50}{\metre}) for the location
             \ang{120.9375}~W, \ang{30.9375} N. Shown are the annual cycles for
             three parameter values $\mathbf{u}_0$ (as ``reference point''),
             $\bar{\mathbf{u}}$ (a neighbouring point) and $\tilde{\mathbf{u}}$
             (a point in a closer vicinity) defined in Table
             \ref{table:ModelParameterValues}. The surrogate is built using
             parameter vector $\mathbf{u}_0$, the reason why the solution of the
             surrogate is omitted at the reference point.}
    \label{fig:N_N-ANN_AnnualCycleSurrogate}
  \end{figure}

  Using the prediction of the ANN as initial concentration for a spin-up over
  \num{50} model years as low-fidelity model (see Section
  \ref{sec:Low-Fidelity-Model-ANNinit}) was reasonable to construct a reliable
  surrogate in conjunction with the multiplicative response correction for the
  SBO using the N model. The annual cycles of phosphate in Figure
  \ref{fig:N_N-ANN_AnnualCycleSurrogate} indicate an almost equivalent
  approximation of the high-fidelity model concentration by the surrogates
  using, on the one hand, the low-fidelity model with initial concentration
  predicted by the ANN and, on the other hand, the low-fidelity model with
  constant initial concentrations. Although the concentrations differed for
  both low-fidelity models, the multiplicative correction technique improved
  the accuracy of the low-fidelity models and resulted in nearly the same
  approximation.

\begin{figure}[!tb]
    \centering
    \subfloat[Cost function value $J\left( \mathbf{y}_f \right)$.]{\includegraphics{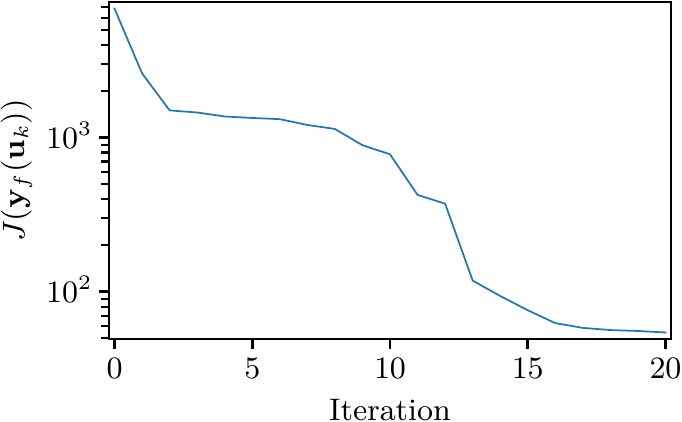}}
    \quad
    \subfloat[Step-size norm of the trust-region radius $\delta_k$.]{\includegraphics{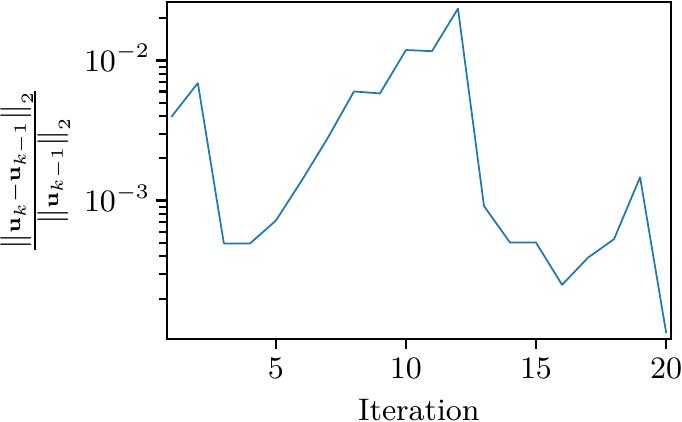}}
    \caption{Convergence of the cost function value and the step-size norm of
             the trust-region radius for the exemplary SBO run with the
             prediction of the ANN as initial concentration for the spin-up as
             low-fidelity model ($\mathbf{y}_c^{\text{ANN-N}}$).}
    \label{fig:N_N-ANN_Convergence}
  \end{figure}

\begin{figure}[!tb]
    \centering
    \subfloat{\includegraphics[width=0.315\textwidth]{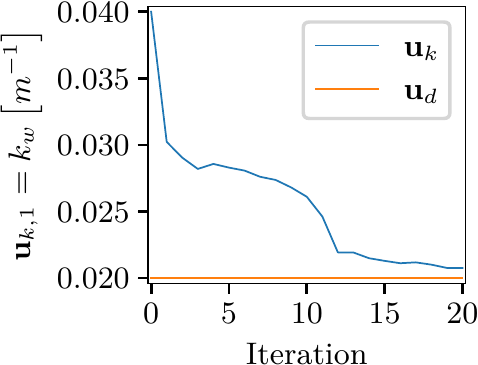}}
    \quad
    \subfloat{\includegraphics[width=0.315\textwidth]{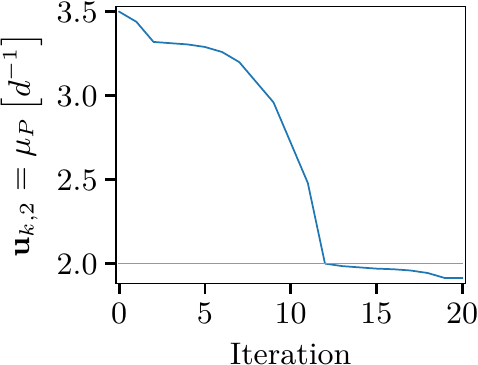}}
    \quad
    \subfloat{\includegraphics[width=0.315\textwidth]{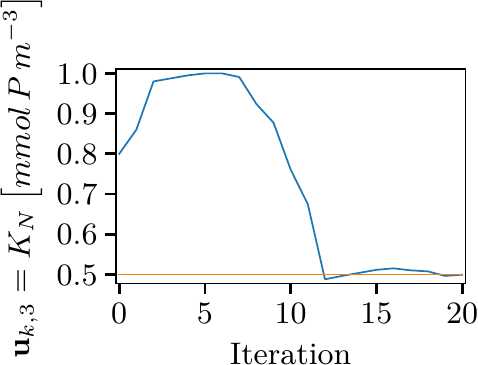}}
    \quad
    \subfloat{\includegraphics[width=0.315\textwidth]{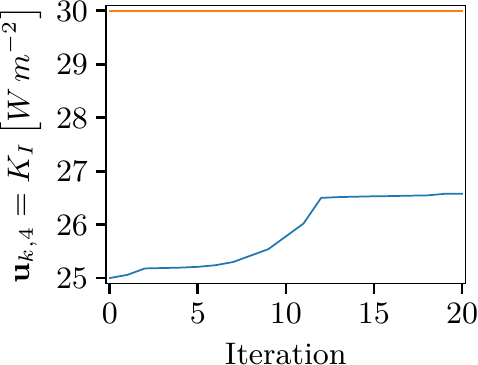}}
    \quad
    \subfloat{\includegraphics[width=0.315\textwidth]{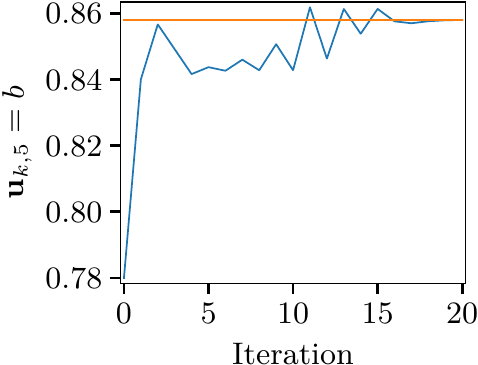}}
    \caption{Convergence of the single parameter values $\mathbf{u}_{k,i}$ for
             each iteration of the exemplary SBO run with the prediction of the
             ANN as initial concentration for the spin-up as
             low-fidelity model ($\mathbf{y}_c^{\text{ANN-N}}$).}
    \label{fig:N_N-ANN_ConvergenceParameter}
  \end{figure}

\begin{figure}[p]
    \centering
    \subfloat[$\mathbf{y}_d$]{\includegraphics[width=0.315\textwidth, viewport=0 2 137 88, clip]{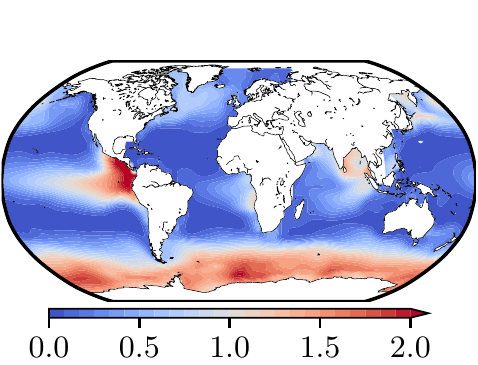}}
    \quad
    \subfloat[$\mathbf{y}_f (\mathbf{u}_0)$]{\includegraphics[width=0.315\textwidth, viewport=0 2 137 88, clip]{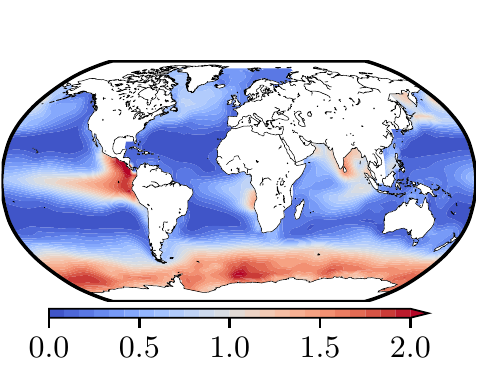}}
    \quad
    \subfloat[$\mathbf{y}_f (\mathbf{u}_5)$]{\includegraphics[width=0.315\textwidth, viewport=0 2 137 88, clip]{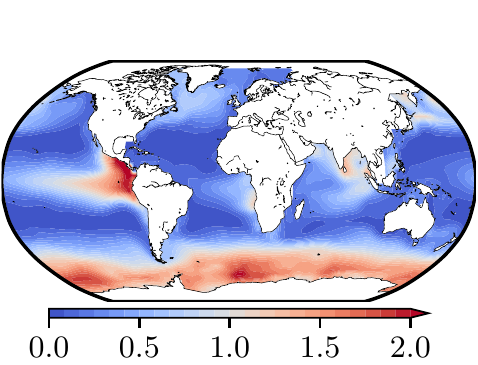}}
    \quad
    \subfloat[$\mathbf{y}_f (\mathbf{u}_{10})$]{\includegraphics[width=0.315\textwidth, viewport=0 2 137 88, clip]{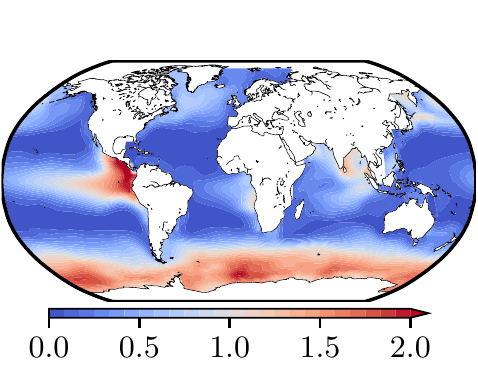}}
    \quad
    \subfloat[$\mathbf{y}_f (\mathbf{u}_{15})$]{\includegraphics[width=0.315\textwidth, viewport=0 2 137 88, clip]{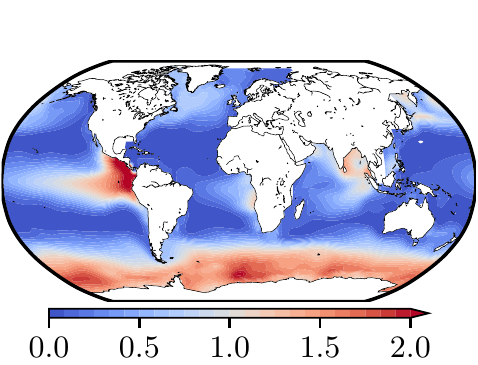}}
    \quad
    \subfloat[$\mathbf{y}_f (\mathbf{u}_{20})$]{\includegraphics[width=0.315\textwidth, viewport=0 2 137 88, clip]{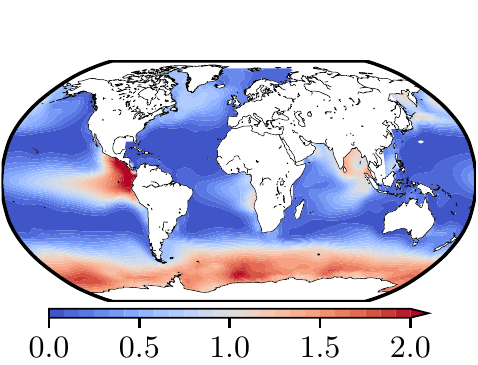}}
    \caption{High-fidelity model output $\mathbf{y}_f$ obtained by the
             exemplary SBO run with the prediction of the ANN as initial
             concentration for the spin-up as low-fidelity model
             ($\mathbf{y}_c^{\text{ANN-N}}$) at the beginning and after
             \num{5}, \num{10}, \num{15} and \num{20} iterations (i.e.,
             evaluated with the parameter vectors $\mathbf{u}_0$,
             $\mathbf{u}_5$, $\mathbf{u}_{10}$, $\mathbf{u}_{15}$ and
             $\mathbf{u}_{20}$) as well as the target data $\mathbf{y}_d$.
             Shown are the phosphate concentrations on the surface layer
             (\SIrange{0}{50}{\metre}) at the first time step of the model
             year (in January).}
    \label{fig:N_N-ANN_Surface}
  \end{figure}

\begin{figure}[p]
    \centering
    \subfloat[Location \ang{120.9375} W, \ang{30.9375} N]{\includegraphics{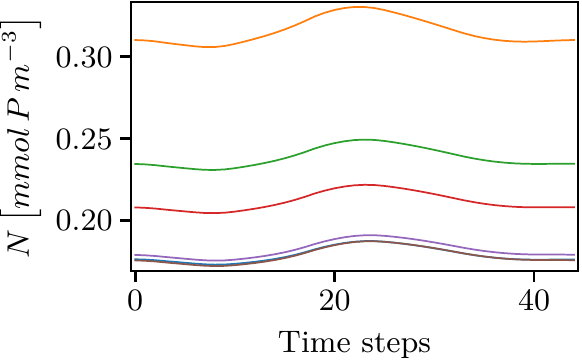}}
    \quad
    \subfloat[Location \ang{90.0} E, \ang{0.0} N]{\includegraphics{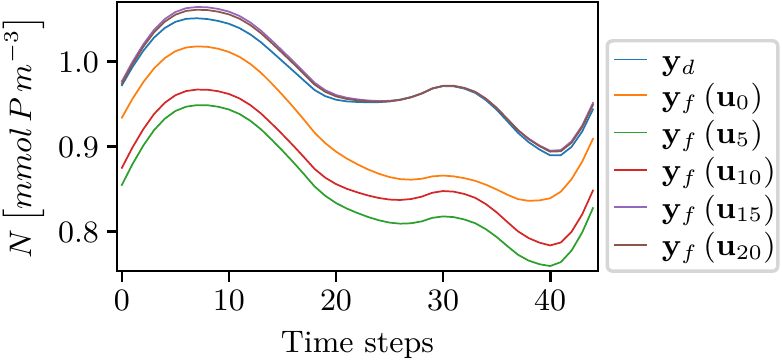}}
    \caption{Annual cycles of phosphate of the high-fidelity model output
             $\mathbf{y}_f$ obtained by the exemplary SBO run with the
             prediction of the ANN as initial concentration for the spin-up as
             low-fidelity model ($\mathbf{y}_c^{\text{ANN-N}}$) at the
             beginning and after \num{5}, \num{10}, \num{15} and \num{20}
             iterations (i.e., evaluated with the parameter vectors
             $\mathbf{u}_0$, $\mathbf{u}_5$, $\mathbf{u}_{10}$,
             $\mathbf{u}_{15}$ and $\mathbf{u}_{20}$) as well as the target
             data $\mathbf{y}_d$ on the surface layer (\SIrange{0}{50}{\metre})
             for two distinct locations.}
    \label{fig:N_N-ANN_AnnualCycle}
  \end{figure}

  The parameter identification using the SBO identified four out of five model
  parameters of the N model. Figure \ref{fig:N_N-ANN_Convergence} shows the
  convergence of the tracer concentrations to the target data $\mathbf{y}_d$
  using different step sizes during the optimization. Except for parameter
  $K_I$, the single parameters converged to the parameter values of the optimal
  parameter vector $\mathbf{u}_d$ (Figure
  \ref{fig:N_N-ANN_ConvergenceParameter}). As a result of a step size less
  than the threshold $\gamma$, the SBO terminated after \num{20} iterations.
  For this optimization run, we applied parameters $m_{\text{incr}} = 2$ and
  $m_{\text{decr}} = 2$ for the update of the trust-region radius. Figures
  \ref{fig:N_N-ANN_Surface} and \ref{fig:N_N-ANN_AnnualCycle} illustrate the
  continuous improvement of the agreement of the high-fidelity solution and the
  target data during the optimization.

  The computational effort of the parameter identification using the SBO was
  low. To obtain parameter vector $\mathbf{u}_{20}$, the SBO required
  \num{35} high-fidelity model evaluations and about \num{8000} evaluations of
  the surrogate. In the exemplary SBO run, the surrogate optimization needed up
  to \num{50} iterations. Indeed, the number of surrogate evaluations could
  be reduced if we restricted the iterations for each surrogate optimization
  to ten. Apart from reducing computational costs, the limitation of the 
  iterations did not lead to an adequate identification of the model parameters
  because the cost function reached a value of \num{1300} only.

  \section{Conclusions}
  \label{sec:Conclusions}

    The SBO is based on a low-fidelity model that, on the one hand, can be
evaluated computationally inexpensive and, on the other hand, is an appropriate
approximation of the high-fidelity model. A suitable correction technique
improves additionally the accuracy of the low-fidelity model building the
surrogate. We have compared three different low-fidelity models in conjunction
with the multiplicative response correction already employed by
\textcite{PPKOS13}. In addition to the truncated spin-up \parencite{PPKOS13},
we applied a neural network only and in combination with the truncated spin-up
as low-fidelity model.

Using the SBO, the parameter optimization yielded similar results for the
low-fidelity models including and excluding the application of the neural
network except for the direct use of the prediction. With the exception of
model parameter $K_I$ modeling the light intensity compensation, the SBO
identified all model parameters for both biogeochemical models and confirmed the
results of \textcite{PPKOS13} for the N-DOP model. However, the parameter
identifications were not optimal. A reason for the mismatch, on the one hand,
was possibly the low accuracy of the underlying low-fidelity model as well as a
weak sensitivity with regard to the model parameters as already observed by
\textcite{RSSSWP10} for a different biogeochemical model. For instance, the
construction of the surrogate considering information about the high-fidelity
model could increase the accuracy. On the other hand, the trust region radius
limited the convergence of parameter $K_I$ because the admissible parameter
range of this parameter was much larger than for the other parameters and, thus,
larger adjustments were necessary.

The use of neural networks to construct the surrogate reduced significantly
the computational costs of a parameter optimization. A reduction of
these costs is desirable because a parameter identification in climate models
is very expensive \parencite{KwoPri06, KwoPri08, MaFeDo12}. Indeed, the SBO,
however, reduces the computational costs compared to standard optimization
routines but the reduction of costs for the low-fidelity model decreases these
further. If a neural network is available, the computational costs of the
prediction are negligible. Therefore, the costs only originated from the
subsequent truncated spin-up, if this was used. However, the prediction of the
ANN alone was not suitable as low-fidelity model because only minimal
concentration changes arose from the prediction using various parameter vectors
in a small environment. A neural network trained with parameter vectors in such
a small environment, nevertheless, had the ability to predict the
concentrations reasonably for a random parameter vector of this environment
whereas the prediction was often insufficient for parameter vectors outside
this environment. The training of a neural network, furthermore, is
computationally costly and depends on the availability of a sufficient amount of
training data. Future work should, therefore, include the training of an ANN
that better predicts the concentrations locally and, thus, serves as
low-fidelity model for the construction of a reliable surrogate with the
multiplicative correction. For this purpose, reinforcement learning
\parencite{SutBar18, MKSRV+15} could help to continuously improve the neural
network during an optimization run. Moreover, further future work should
include the application of the SBO for parameter identification using a
normalization of the admissible parameter range for each component of the
parameter vector as well as using real measurement data that, for example, is
necessary for model assessment \parencite[see e.g.,][]{KrKhOs10} and model
calibration \parencite[see e.g.,][]{KSKSO17, Kri17}.

In summary, the main points of this paper are the following:
\begin{itemize}
  \item Using the truncated spin-up as a low-fidelity model, the SBO was a
        computationally efficient method for the parameter identification of
        marine ecosystem models.
  \item The prediction of the presented ANN as low-fidelity model was unsuitable
        for an SBO.
  \item Using the prediction of the ANN as an initial concentration for a
        truncated spin-up, the low-fidelity model represented an acceptable
        alternative to the truncated spin-up with slightly higher computational
        costs.
\end{itemize}

  \section*{Code and data availability}
  
    The code used to generate the data in this publication  is available at
\url{https://github.com/slawig/bgc-ann} and \url{https://metos3d.github.io/}.
All used and generated data are available at
\url{https://doi.org/10.5281/zenodo.5643667} \parencite{PfeSla21bData}.

\printbibliography

\end{document}